\documentclass[graybox,vecarrow]{svmult}

\usepackage{mathptmx}       % selects Times Roman as basic font
\usepackage{helvet}         % selects Helvetica as sans-serif font
\usepackage{courier}        % selects Courier as typewriter font
\usepackage{type1cm}        % activate if the above 3 fonts are
\usepackage{makeidx}         % allows index generation
\usepackage{graphicx}        % standard LaTeX graphics tool
\usepackage{multicol}        % used for the two-column index
\usepackage[bottom]{footmisc}% places footnotes at page bottom
\makeindex             % used for the subject index
\newcommand{\sto}{SrTiO$_3$}
\newcommand{\lao}{LaAlO$_3$}
\newcommand{\lto}{LaTiO$_3$}
\newcommand{\tio}{TiO$_2$}
\newcommand{\kto}{KTaO$_3$}
\newcommand{\abo}{ABO$_3$}
\newcommand{\tg}{$t_{2g}$}

\newcommand{\dxy}{$d_{xy}$}

\newcommand{\dxzyz}{$d_{xz/yz}$}

\newcommand{\figref}[1]{Fig.~\ref{#1}}

\newcommand{\secref}[1]{Sect.~\ref{#1}}

\begin{document}

\title*{ARPES Studies of Two-Dimensional Electron Gases at Transition Metal Oxide Surfaces}
\author{Siobhan McKeown Walker and Flavio Y. Bruno and Felix Baumberger}
\institute{Siobhan McKeown Walker \at University of Geneva, 24 Quai Ernest-Ansermet, Geneva, CH-1211, Switzerland \email{siobhan.mckeown@unige.ch}
\and Flavio Y. Bruno \at University of Geneva, 24 Quai Ernest-Ansermet, Geneva, CH-1211, Switzerland \email{flavio.bruno@unige.ch}
\and Felix Baumberger \at University of Geneva, 24 Quai Ernest-Ansermet, Geneva, CH-1211, Switzerland and Swiss Light Source, Paul Scherrer Institut, Villigen, CH-5232, Switzerland \email{felix.baumberger@unige.ch}}
%University of Geneva, 24 Quai Ernest-Ansermet, Geneva, CH-1211, Switzerland and Paul Scherrer Institut,
\maketitle

\abstract{High mobility two-dimensional electron gases (2DEGs) underpin today's silicon based devices and are of fundamental importance for the emerging field of oxide electronics. Such 2DEGs are usually created by engineering band offsets and charge transfer at heterointerfaces. However, in 2011 it was shown that highly itinerant 2DEGs can also be induced at bare surfaces of different transition metal oxides where they are far more accessible to high resolution angle resolved photoemission (ARPES) experiments. Here we review work from this nascent field which has led to a systematic understanding of the subband structure arising from quantum confinement of highly anisotropic transition metal $d$-states along different crystallographic directions. We further discuss the role of different surface preparations and the origin of surface 2DEGs, the understanding of which has permitted control over 2DEG carrier densities. Finally, we discuss signatures of strong many-body interactions and how spectroscopic data from surface 2DEGs may be related to the transport properties of interface 2DEGs in the same host materials.}

\section{Introduction}

Oxide surfaces and interfaces can host electronic states that differ from those in the bulk. This offers new possibilities for electronic structure design and has motivated an increasing number of studies investigating epitaxial heterostructures. The \abo{} perovskite transition metal oxides (TMOs) have received much attention in this context because their quasi-cubic structures and compatible lattice constants make them well suited to heteroepitaxy growth~\cite{Zubko2011}. Moreover, they show diverse bulk properties including ferro- and antiferromagnetism, ferroelectricity or superconductivity. These phases are largely controlled by the occupation of the transition metal $d$-shell and by subtle changes in bond angles, rendering \abo{} TMOs suitable to electronic structure engineering in heterostructures by exploiting interfacial charge transfer, strain and octahedral tilting patterns ~\cite{Tokura1993,Mannhart2010,Zubko2011}. 

Of particular interest are high-mobility two-dimensional electron gases (2DEGs) in \abo{} perovskites. These systems have the potential to underpin a new generation of oxide electronics by exploiting not only the various phases of the parent oxide as carrier densities are tuned, but also phases and properties unique to the oxide surface or interface 2DEG~\cite{Mannhart2010,Jany2014}. Determining the intricacies of the electronic band structure of such TMO 2DEGs is an important step towards understanding the underlying physics of these systems and facilitates the engineering of desirable properties. However, the intrinsically buried nature of interface 2DEGs poses substantial experimental challenges. High-resolution angle resolved photoemission (ARPES) using UV excitation, a standard technique for band structure determination, has insufficient probing depth to study important systems such as the \lao/\sto{} (LAO/STO) interface 2DEG, despite them being buried beneath only a few unit cells. %The standard band structure probe of high-resolution angle resolved photoemission (ARPES) using UV excitation has insufficient probing depth to study important systems such as the \lao/\sto{} (LAO/STO) interface 2DEG, despite them being buried underneath only a few unit cells. 
This restricts photoemission studies of interface 2DEGs to the soft or hard X-ray regime where the effective resolution in energy and momentum is reduced. Other spectroscopic techniques that were successfully applied to oxide interfaces such as X-ray absorption (XAS)~\cite{Salluzzo2009} or resonant inelastic X-ray scattering (RIXS)~\cite{Berner2010,Zhou2011} give valuable information on orbital symmetries and collective excitations but do not offer direct momentum space resolution. Microscopic electronic structure information from interfaces has also been deduced from quantum oscillation data. This technique is exceptionally precise but requires very high mobilities, which are hard to achieve in correlated electron systems, and the data is often difficult to interpret~\cite{Fete2012,Moetakef2012,McCollam2014,Kim2011}. These challenges have motivated an alternative approach to the creation and spectroscopic investigation of oxide 2DEGs. Recognizing that the fundamental electronic properties of 2DEGs are defined by their host material and the electrostatic boundary conditions, in 2011 Meevasana \textit{et al.}~\cite{Meevasana2011a} and Santander-Syro \textit{et al.}~\cite{Santander-Syro2011a} reported that a 2DEG showing hallmarks of the band structure predicted for the LAO/STO interface can be created on the bare (001) surface of \sto{} (STO) where it is accessible to high-resolution ARPES experiments. This approach has subsequently been extended to different TMO host materials and surface orientations revealing the fundamental electronic properties of oxide 2DEGs and a common framework for describing system-to-system variation. In this chapter, we review the present status of this emerging field.

\subsection{Metallic subbands at the surface of an insulator}

\begin{figure}
	\centering
	\includegraphics[width =0.9\textwidth]{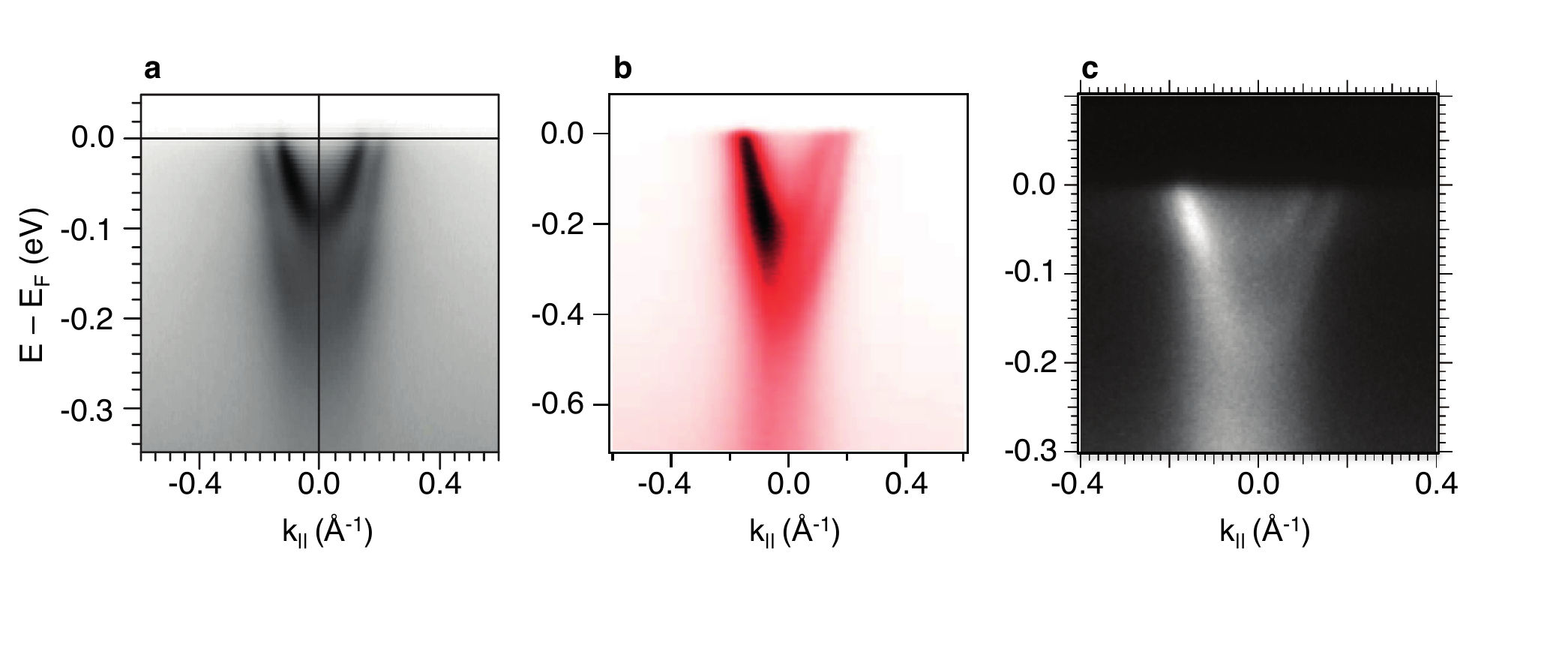}
	\caption{ Energy-momentum dispersions of surface 2DEGs measured by VUV ARPES on the (001) surfaces of  (a) SrTiO$_3$, (b) KTaO$_3$ and (c) anatase \tio{}. High intensity (black, black and white in (a), (b) and (c) respectively) delineates the electronic states of the charge accumulation layer. Adapted from Santander-Syro \textit{et al.} \cite{Santander-Syro2011a}, King \textit{et al.} \cite{King2012a} and R\"odel \textit{et al.} \cite{Rodel2015} respectively. 		
	 }
	\label{2DEG_F1}  
\end{figure}

We will focus the discussion on ARPES studies of 2DEGs with $d$ orbital character in the transition metal oxides \sto~\cite{Santander-Syro2011a,Meevasana2011a}, \kto{} (KTO)~\cite{King2012a,Santander-Syro2012,Bareille2014b} and anatase \tio~\cite{Moser2013a,Rodel2015}. We note that electron accumulation layers have also been observed on the surface of the transparent conducting oxides CdO~\cite{Piper2008,King2010} and In$_2$O$_3$~\cite{Zhang2013} but these 2DEGs derive from free-electron like $s$ states and will not be discussed here.
In stoichiometric form, STO, KTO and anatase \tio{} are band insulators with a $d^0$ configuration of the transition metal ion. Importantly, all three materials are susceptible to chemical doping \cite{Schooley1965,Spinelli2010,Wemple1965,Uwe1979,Forro1994,Jacimovic2013} which introduces electrons into the conduction band minimum resulting in a three dimensional bulk metallic state. Using appropriate surface preparations, electron accumulation layers that are independent of the residual bulk doping have been reported in all three of these TMOs. As shown in \figref{2DEG_F1}, these charge accumulation layers all show multiple subbands, which is a key-signature of quantum confinement and is adopted as the finger-print of a 2DEG in ARPES measurements  throughout this review.
 % that lift the orbital degeneracies found in the bulk.
%  \textit{This is a key-signature of quantum confinement and will be adopted as the defining property of a 2DEG throughout this review.}
In the following we will discuss the subband structures of 2DEGs in STO, KTO and anatase \tio{} in detail, exploring not only material dependent electronic properties, but also the influence of the crystallographic orientation of the surface on 2DEG characteristics. We will further discuss the origin of these metallic states on the surfaces of insulating TMOs and briefly review different approaches for calculating their band structure. We will also summarize very recent ARPES studies that provide insight into the nature of many-body interactions in oxide 2DEGs.

\section{Origin of surface 2DEGs in TMOs}
A 2DEG arises as the conduction band minimum of an insulation or semiconducting crystal is dragged below the chemical potential by an electrostatic potential over a narrow region. Understanding the origin of the corresponding electric field in 2DEG systems is an important step towards achieving carrier density control, which in-turn underlies device functionality. While in conventional semiconductors it is well established that the potential gradient arises from workfunction mis-match, the origin of both the attractive confining potential and the excess charge carriers in TMO surface and interface systems is more ambiguous. For example the origin of the native charge carriers at the LAO/STO interface is still actively debated and consequently systematic control of their density has remained elusive.

The first two publications that reported the STO (001) surface 2DEG, suggested that the origin of both the electrostatic band bending and charge carriers may be surface oxygen vacancies (OVs) \cite{Santander-Syro2011a,Meevasana2011a}. In this scenario two excess electrons are released as a positively charged  OV is created. This positive charge at the STO surface must be screened by the excess electrons which can form either localized states near the vacancy, or an itinerant accumulation layer which manifests as the observed 2DEG. Additionally the authors of reference \cite{Meevasana2011a} observed that the 2DEG bandwidth and density increase as the surface is irradiated with synchrotron light and remains constant in UHV conditions when not irradiated, leading to the hypothesis that the STO surface 2DEG originates from light-induced oxygen vacancies. In the following we will describe the experimental evidence for this scenario. 

\subsection{UV Induced Oxygen Vacancies}
\label{OV_section}

\begin{figure}
  		\centering
    \includegraphics[width=0.7\textwidth]{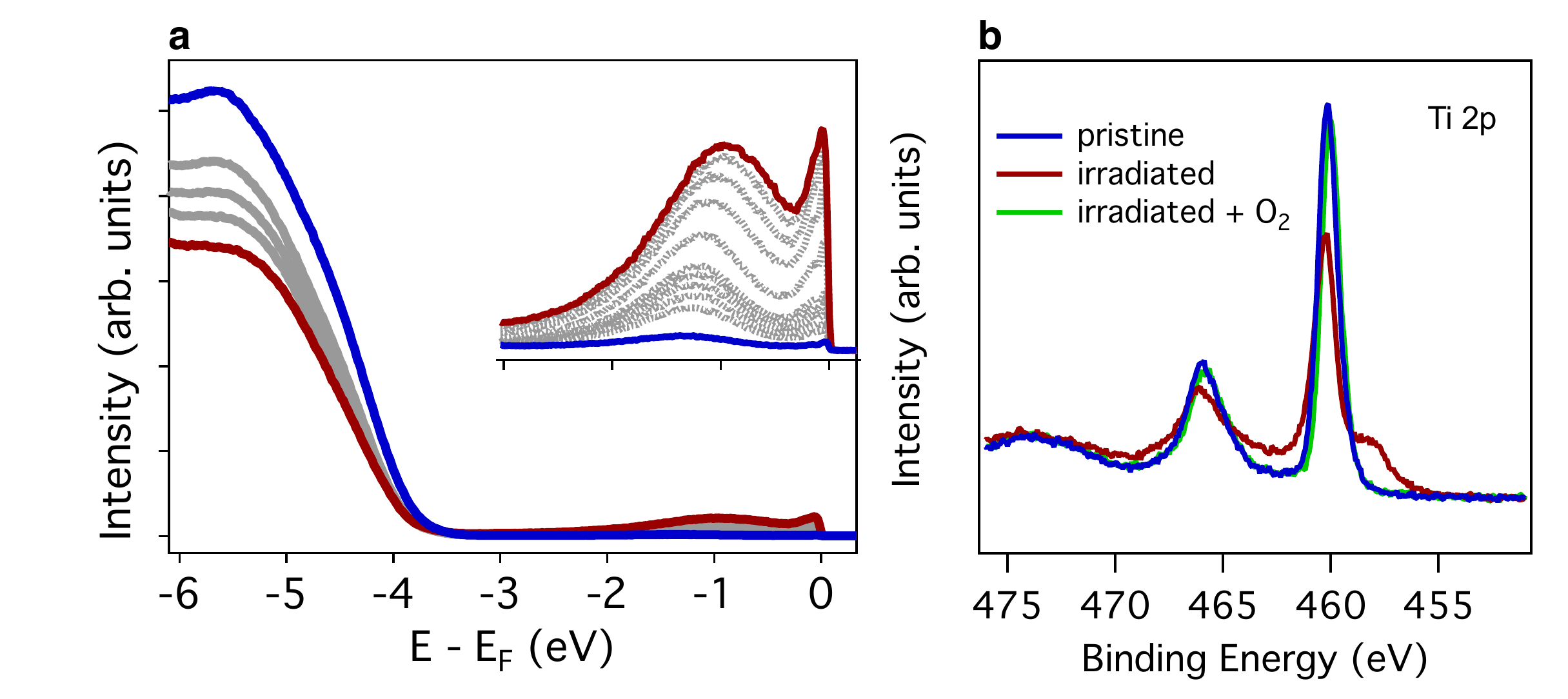}
  	\caption{(a) Angle integrated energy distribution curves showing the valence band, in-gap state and 2DEG evolution on the fractured surface of La:STO (001) after negligible (blue) or prolonged (red) UV irradiation and for intermediate irradiation times (gray) (unpublished). (b) XPS showing the Ti$^{3+}$ core level under the same conditions and after exposure to 0.5 Langmuir of O$_2$ (green), adapted from McKeown Walker \textit{et al.} \cite{McKeown2015}.}
  	\label{VB_Ti2p}
  \end{figure}

 Both Santander-Syro \textit{et al.} \cite{Santander-Syro2011a} and Meevasana \textit{et al.} \cite{Meevasana2011a} saw evidence for band bending at the (001) STO surface. They measured angle integrated energy distribution curves (EDCs), similar to those in \figref{VB_Ti2p}(a) which shows the O2$p$ valence band (VB) whose maximum is around 4 eV below the Fermi level before significant irradiation (blue), after long UV irradiation (red) and at intermediate times (grey). Its can be seen that the VB appears to shift to higher binding energies as the surface is irradiated. Together with small core level shifts observed by X-ray photoemission spectroscopy (XPS) \cite{Dudy2016,Plumb2014} this suggests the presence of downward band-bending at the surface induced by the UV radiation \cite{Meevasana2011a,McKeown2015,McKeown2014,Dudy2016}. This band bending appears concomitant with the 2DEG peak at the Fermi level (see inset) and its magnitude, which can be broadly quantified by the shift of the VB leading edge mid-point, should be related to the 2DEG bandwidth. However, the exact magnitude of the surface band bending is difficult to extract from such spectra since they represent an average of the energy shift in each unit cell over the photoemission probing depth, and the form of the VB also evolves as the surface is irradiated.
 
In addition references \cite{Santander-Syro2011a} and \cite{Meevasana2011a} observed a non-dispersive in-gap (IG) state approximately 1.3 eV below the Fermi level that grows in intensity with increasing irradiation time (see inset of \figref{VB_Ti2p}(a)). Previous photoemission measurements on reduced STO, Nb:STO and La:STO also observed this IG state at the STO surface and associated it with oxygen vacancy defects states \cite{Henrich1978,Aiura2002}. Further evidence that electrons localized on or near oxygen vacancy sites would form such an IG state was provided by DFT calculations of STO with oxygen deficient surfaces \cite{Jeschke2015}.

 \begin{figure}
 	\centering
 	\includegraphics[width=0.8\textwidth]{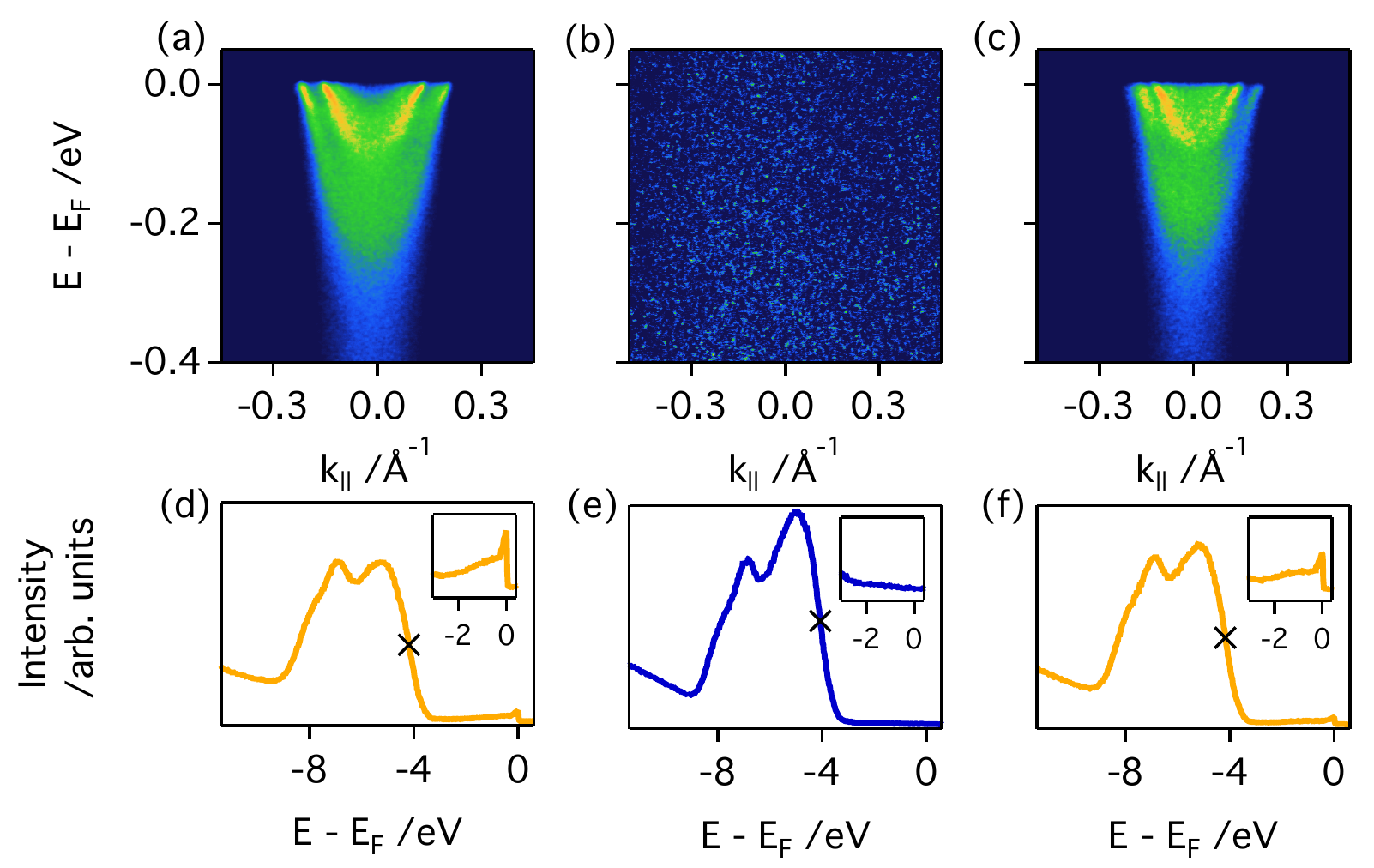}
 	\caption{Creation and annihilationof the 2DEG at the STO(001) surface. (a) Dispersion plot of the high density STO(001) surface 2DEG after initial irradiation with 52 eV photons. (b,c) The 2DEG disappears after in-situ exposure to 0.5 Langmuir of O$_2$ and reappears following further irradiation with 52 eV photons. (d-f) Angle integrated valence band spectra corresponding to the states in (a-c). The magnified insets show the intensity at the Fermi level. The valence band leading edge midpoint (VB LEM) is marked by a cross. All data were measured in the second Brillouin zone with 28 eV, $s-$polarized light. From McKeown Walker \textit{et al.} \cite{McKeown2015}.}
 	\label{am_f2}
 \end{figure}
%It was e.

McKeown Walker \textit{et al.}~\cite{McKeown2015} demonstrated that, as shown in \figref{am_f2}(a)-(c), introducing extremely low doses of molecular oxygen into the UHV chamber eliminates the surface 2DEG and that subsequent irradiation with UV light causes the 2DEG density to recover. This explicitly demonstrated that the 2DEG is an accumulation layer of electrons at the surface of STO screening positive charge resulting from light induced oxygen vacancies. This experiment also confirmed that the IG state is associated with light-induced OVs providing evidence that these defects at the STO surface result in both itinerant and localized electronic states. Reference \cite{McKeown2015} also demonstrate that the efficiency with which OV are created depends on the photon energy, suggesting that the dominant mechanism by which OV are created is inter-atomic core-hole Auger decay \cite{Knotek1978,Knotek1984}. Using this knowledge McKeown Walker \textit{et al.} succeeded in controlling the density of the STO 2DEG by either measuring at a photon energy below the threshold for efficient OV creation or by measuring at higher photon energies while maintaining a partial pressure of oxygen during the measurement.

 The bandwidth and density of the STO surface 2DEG do not grow indefinitely under synchrotron radiation. After a finite irradiation period the band width saturates. However, as reported by Dudy \textit{et al.}~\cite{Dudy2016} the intensity of both the IG state and 2DEG do not saturate over the same time scale. These authors suggest that the dynamic equilibrium between OV creation and annihilation at finite oxygen partial pressure leads to OV clusters and electronic phase separation. Alternative explanations include the migration of ions within the lattice due to high electric fields at the surface, and that the ratio of localized and itinerant electrons donated by an oxygen vacancy evolves as a function of OV density due to a changes in the balance of correlations \cite{Lin2013}.

 Qualitatively the same UV sensitive behaviour of the two-dimensional electron gas density, VB and core level shifts and IG state intensity has been observed for various low-index surfaces of STO \cite{McKeown2014,Wang2014a,Rodel2014} and for the surface of anatase \tio{} \cite{Rodel2015}. The carrier density of states observed by ARPES at the surface of anatase \tio{} could be controlled by tuning the dynamic equilibrium between OV creation by UV light and re-oxidation due to finite oxygen partial pressure in the chamber \cite{Moser2013a}. Indeed the role of oxygen vacancies in \tio{} thin films is even more dramatic with excessive irradiation leading to a local destruction of the 2DEG state \cite{Rodel2015}. Implicit in these observations is that reconstructions of the crystal surfaces \cite{Plumb2014,Chen2015,Delugas2015} do not dominate surface charge accumulation in STO or \tio. This is true even in the case of (111) and (110) surfaces of STO which are polar. On the other hand, for the 2DEG at the strongly polar (001) surface of \kto{} only a weak evolution of the 2DEG bandwidth is seen as the surface is irradiated \cite{King2012a,Santander-Syro2012}, suggesting that defects intrinsic to the cleaved surface may induce a 2DE in this case.
\section{2DEG Subband Structure at the (001), (110) and (111) Surfaces of \sto}
\label{bandstructure}
In this section we will describe the subband structures of STO 2DEGs observed by ARPES in more detail. We will demonstrate that the characteristic features of the electronic structure can be understood on a qualitative level in terms of quantum confinement thereby justifying the identification of these states as two dimensional electron gases. We will also discuss the role of crystallographic orientation of the surface in modulating the effects of quantum confinement on the 2DEG band structure.

\subsection{\sto{} (001)}
\label{STO_001}
\begin{figure}
	\centering
	\includegraphics[width =0.8 \textwidth]{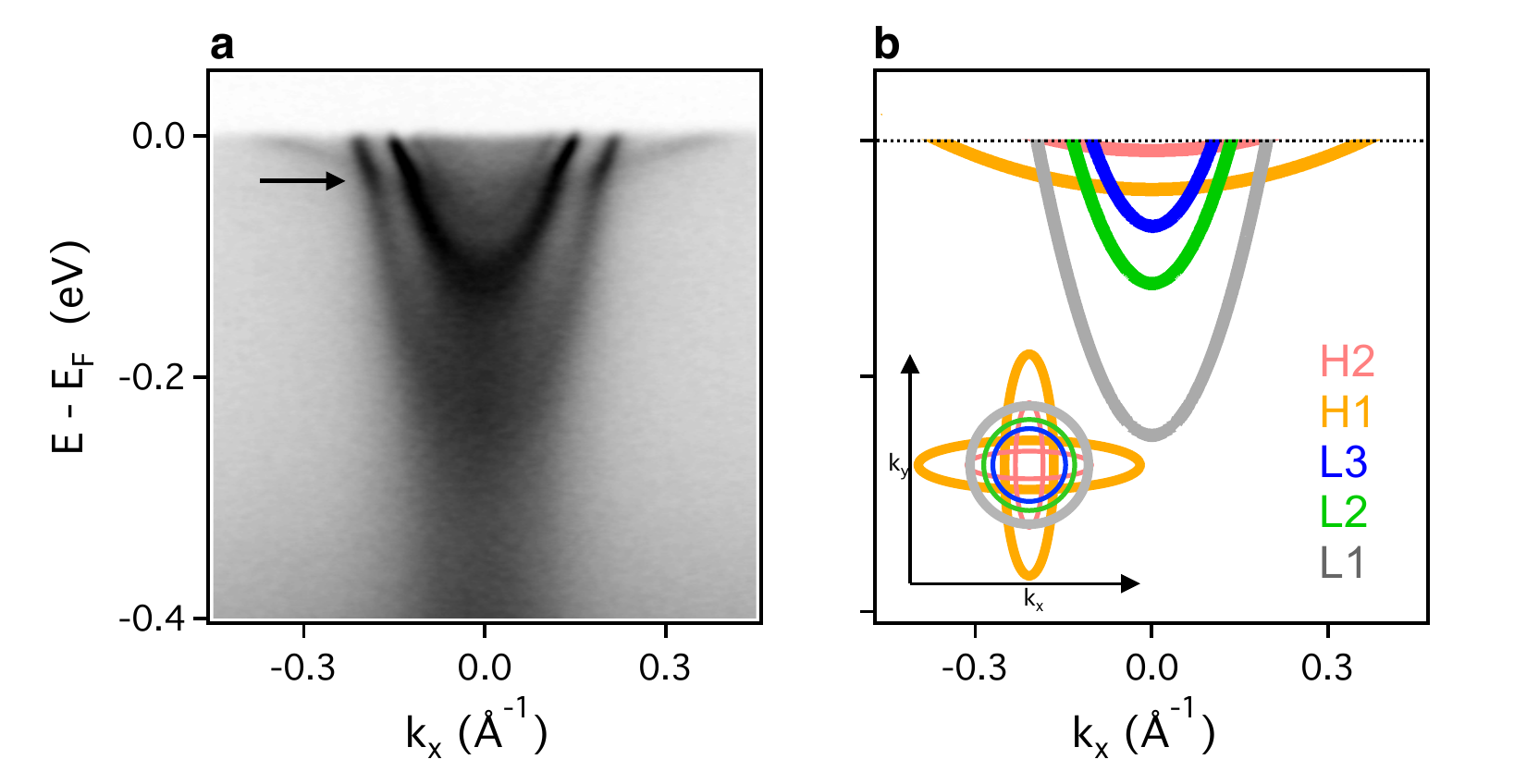}
	\caption{Electronic structure of the 2DEG at the (001) surface of \sto{}. (a) Energy momentum dispersion plot measured in the second Brillouin zone. Data shown are the sum of measurements taken with s- and p- polarized light at 52 eV photon energy at 10 K. The arrow indicates the kink in the disperion of the L1, L2 and L3 subbands due to electron-phonon interaction. (b) Schematic of the subbands resolved in the ARPES measurements of (a). The three light bands (L1, L2, L3) and the two heavy bands (H1 and H2) are sketched in grey, green, blue, orange and coral respectively. L1, L2 and L3 have predominantly \dxy-orbital character and H1,H2 have predominantly \dxzyz-orbital character. The inset shows a sketch of the corresponding Fermi surface using the same colour scheme to distinguish between the Fermi surface sheets. Adapted from McKeown Walker \textit{et al.} \cite{McKeown2015}.}
	\label{McK2015_F1}
\end{figure}

\figref{McK2015_F1}(a) shows the energy-momentum subband dispersion of the 2DEG measured on the fractured (001) surface of STO. This data, reproduced from reference \cite{McKeown2015}, resolves five subbands, as shown schematically in \figref{McK2015_F1}(b). Three subbands (L1-L3) are highly dispersive indicating a light effective mass for some of the carriers while the two shallowest subbands (H1-H2) resolved in the experiment are much less dispersive. From parabolic fits of the overall dispersion of individual subbands we estimate effective masses of $\sim0.6~m_e$ for L1-L3 and $9-15~m_e$ for (H1-H2) \cite{King2014}. These values are comparable to DFT bulk band masses of the STO conduction band~\cite{VanMechelen2008,VanDerMarel2011} for L1-L3 while they are significantly higher than the heaviest bulk masses for H1-H2.King \textit{et al.}~\cite{King2014} showed that this disparity between the light and heavy subbands arises due the electron-phonon interaction. For the light subbands, which have a bandwidth exceeding the Debye frequency, the dominant effect of electron-phonon interaction is an additional low-energy renormalization of the dispersion clearly discernible in the form of a kink in the subband dispersion, indicated by an arrow in Fig.~\ref{McK2015_F1}. On the other hand, the shallow heavy bands are close to the anti-adiabatic limit and are thus subject to an overall renormalization of the entire occupied bandwidth, which causes the increased overall effective masses.

We have already discussed that these states are induced at the crystal surface and therefore cannot be considered bulk bands. This is also evident from the observation of 5 clearly non-degenerate states at the $\Gamma$ point while the conduction band minimum of bulk STO is formed by only three approximately degenerate bands. The formation of multiple subbands, as well as the higher relative energy of the heavy subbands, are hallmarks of quantum confinement of the 3D bulk conduction band near the surface. 
Therefore we associate the indices 1,2 and 3 of L1-L3 and H1,H2 with the principal quantum numbers of individual quantum well states.
The three light subbands all correspond to circular Fermi surface sheets while the heavy subbands form the long axes of cigar-shaped Fermi surface sheets, as sketched in the inset of \figref{McK2015_F1}(b)~\cite{Santander-Syro2011a,King2012a}. 
By considering the shape of the Fermi surface sheets, the spatial anisotropy of the \tg{} orbitals that form the conduction band of STO and the polarization dependence of the photoemission matrix elements for \tg{} states, it was shown that the light subbands are formed by electrons in \dxy{} orbitals, while the heavy bands have \dxzyz{} orbital character~\cite{Santander-Syro2011a,King2014,Rodel2016}. The surface 2DEG has thus the same orbital ordering found for the LAO/STO interface 2DEG. As described in \secref{models} this is a natural consequence of quantum confinement along the [100] surface normal.

\begin{figure}
 	\sidecaption[t]
  	\includegraphics[width =0.5 \textwidth]{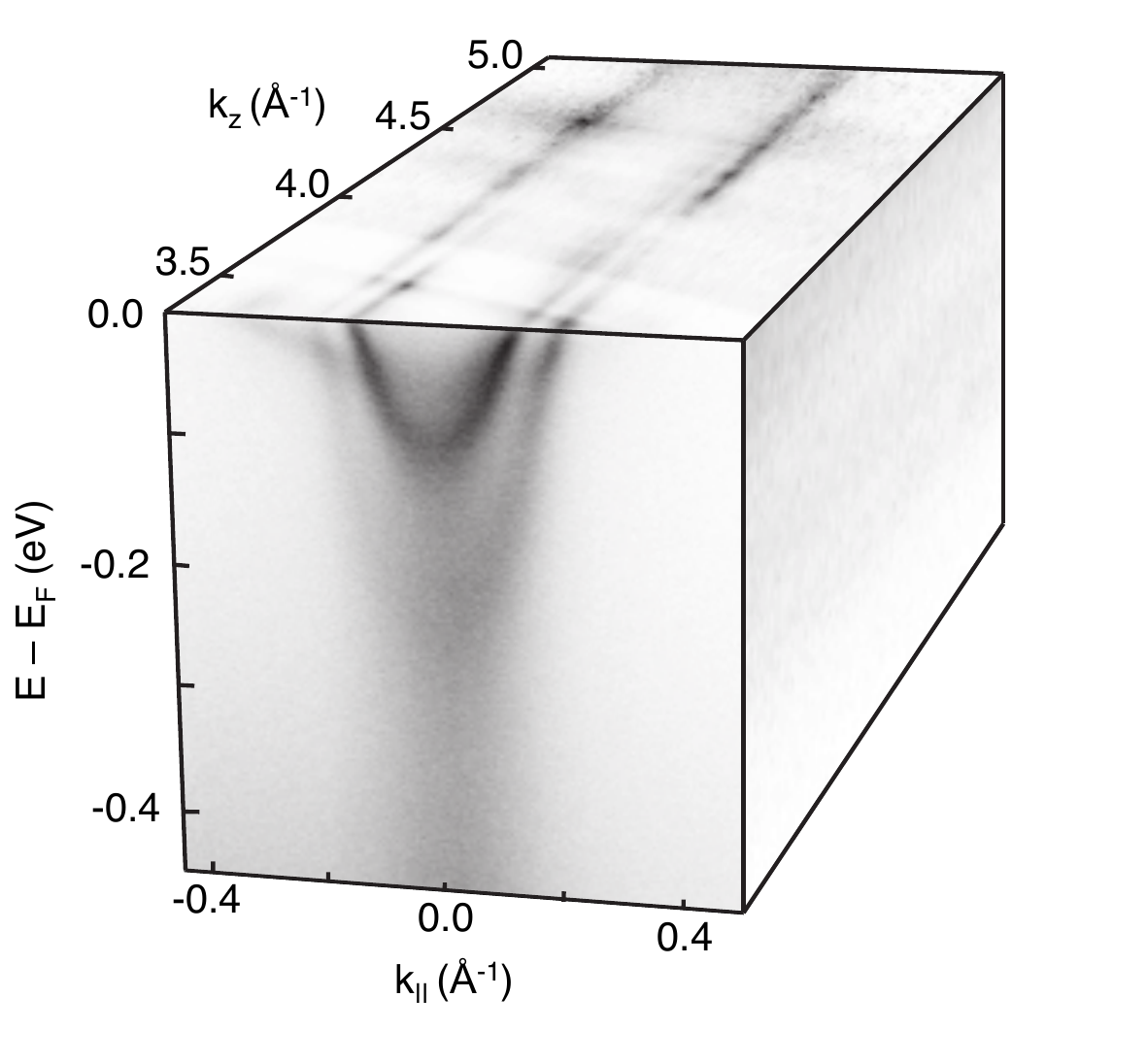}
 	\caption{Two dimensionality of the STO (001) 2DEG. Top Panel: Constant energy map at $E_F$ in the $k_xk_z$ plane. The $k_z$ range shown is approximately one Brillouin zone perpendicular to the sample surface. The non-dispersive nature of the Fermi wave vectors along $k_z$ is indicative of the two dimensional nature of the state. Front Panel: Energy-momentum subband dispersion of the 2DEG. Adapted from Wang \textit{et al.} \cite{Wang2016}.}
 	\label{Wang2016_SF2}
 \end{figure}

Quantum well states are by definition two-dimensional and do not disperse along the confinement direction. Therefore, measuring the subband dispersion along $k_z$ is a direct test for the presence of quantum confinement. 
Experimentally this is achieved by varying the photon energy of the exciting radiation in order to probe the band dispersion perpendicular to the crystal surface. 
One such measurement for the STO (001) 2DEG from Ref.~\cite{Wang2016} is shown in \figref{Wang2016_SF2}.  
The top panel of the data-cube shows the Fermi surface in the $k_xk_z$ plane over a full Brillouin zone. The Fermi wave vectors for the first two light subbands are clearly resolved over an extended range of $k_z$ and are found to be constant within the accuracy of the experiment. This non-dispersive behaviour along the $k_z$ axis confirms the two-dimensional nature of the \dxy{} subbands already inferred in Refs. \cite{Meevasana2011a,Santander-Syro2011a}. 
For the heavy \dxzyz{} subbands the situation is less clear. These bands have generally lower spectral weight which is strongly suppressed away from the bulk $\Gamma$ points making it more difficult to trace their dispersion over an extended range in $k_z$ in order to unambiguously determine their dimensionality.  
Santander-Syro~\textit{et al.}~\cite{Santander-Syro2011a} reported that the first heavy subband H1 is non-dispersive along $k_z$ and thus two-dimensional, while Plumb~\textit{et al.}~\cite{Plumb2014} described H1 as largely three-dimensional. The dimensionality of H2 has not been investigated in the literature to date. An independent argument for a strict two-dimensionality of H1 and H2 comes from their binding energies. As shown in \secref{models} the degeneracy lifting at the $\Gamma$ point between L1 and H1 as well as between H1 and H2 follow naturally from quantum confinement of the bulk conduction band while alternative interpretations are not evident.

The spectral weight distribution of 2D states along $k_z$ is related to the Fourier transform of the real space wave function convolved with the $k_z$ distribution of the photoelectron wave function. Consequently, the spectral weight distribution of 2D states along $k_z$ encodes information about the real space extent of their wavefunctions. Therefore, from the strong intensity of L1 over a full Brillouin zone in $k_z$ we can infer that this state is mostly confined to a single unit cell along the surface normal, while the limited and periodic intensity distribution of H1 indicates a more spatially extended wave function. This is in qualitative agreement with the tight-binding supercell model presented in \secref{models}, even if the complicated oscillations of the photoemission matrix elements have so far prohibited explicit determination of the spatial extent of the wave functions of individual quantum well states.

%This is in semi-quantitative agreement with the tight-binding supercell model presented in \secref{models}, even if the complex oscillations of the photoemission matrix elements have so far prohibited a fully quantitiative determination of the spatial extent of individual quantum well state wave functions.

The STO (001) surface 2DEG with a saturated bandwidth of $\sim250$ meV described in this section has a large carrier density. Evaluating the Luttinger volume of the Fermi surfaces of L1-L3 and H1 for the 2DEG shown in \figref{McK2015_F1} gives $n_{2D}\sim2\cdot10^{14}$ $cm^{-2}$. There is, however, a significant uncertainty in this value arising from higher-order subbands with small volumes and low spectral weight that are not included in this estimate. Moreover, many measurements reported in the literature resolved fewer subbands and correspondingly quote lower values for $n_{2D}$ for the same occupied bandwidth of L1. Despite this uncertainty, the saturated carrier density of the STO (001) surface 2DEG is clearly higher than the sheet carrier  densities reported for the LAO/STO interface 2DEG of typically $n_{2D} = 0.5 - 5\cdot10^{13}$~cm$^{-2}$ \cite{Fete2015} and approaches the value of 0.5 electrons per unit cell $(3.3\cdot10^{14}$~cm$^{-2})$ found in the ideal polar catastrophe scenario for the LAO/STO interface.% A generally lower carrier density at the LAO/STO interface can also be inferred from the occupied bandwidth of $\approx30$~meV found in tunneling spectroscopy of LAO/STO samples compared with the $\approx250$~meV bandwidth of the surface 2DEG \cite{Richter2013}.% \textit{GTO/STO?}

\subsubsection{Universality of TMO surface 2DEG band structure} 
\begin{figure}
 	\centering
 	 	\includegraphics[width =0.95\textwidth]{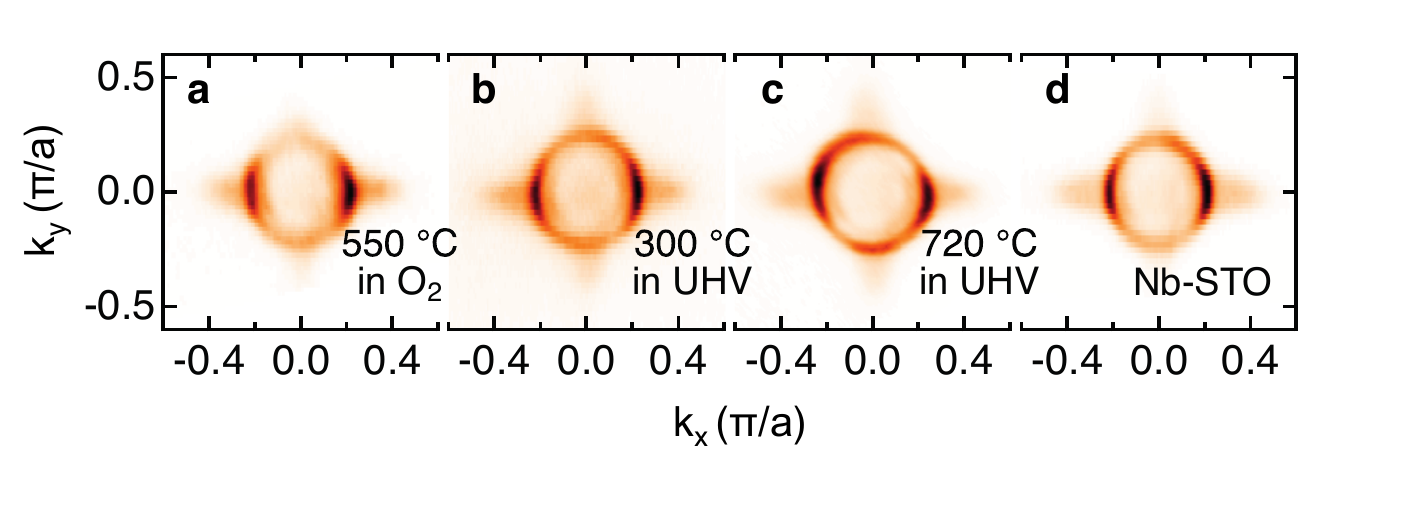}
 	\caption{The Fermi surface of the STO (001) 2DEG in substrates of stoichiometric STO annealed at (a) $550^\circ$C in 100 mbar O$_2$ for 2 hours, (b) $300^\circ$C in UHV for 15 hours (c) $720^\circ$C in UHV for 1 hour and (d) Nb:STO annealed at $550^\circ$C in 100 mbar O$_2$ for 2 hours. The form of the Fermi surface is insensitive to the annealing conditions and bulk doping and is the same as observed on cleaved STO surfaces. This demonstrates the universality of the STO (001) surface 2DEG. From Plumb \textit{et al.} \cite{Plumb2014}.}
 	\label{NCP2014_F2}    
 \end{figure}

% all give the same thing - 2DEG - Ti3+ signal
%various wafer techniques - \cite{Plumb2014}\cite{Chen2015}
%The electronic structure of the STO (001) 2DEG is remarkably insensitive to the surface preparation and the origin and concentration of residual bulk dopants. 
The first publications of the field \cite{Santander-Syro2011a,Meevasana2011a} showed that a 2DEG with virtually identical band structure and density is observed on the bare fractured surfaces of La:STO, stoichiometric STO and reduced STO single crystals with bulk carrier densities up to $n_{3D}\sim1\cdot10^{20}$ cm$^{-3}$. Subsequently, it was shown that the 2DEG can also be induced on \tio{} terminated wafers obtained by mechanical polishing,  \textit{ex situ} etching and different \textit{in-situ} surface preparations~\cite{Wang2016,Plumb2014,Santander-Syro2014,Chen2015}.
\figref{NCP2014_F2} shows the Fermi surface of the STO (001) 2DEG observed on \tio-terminated wafers following different \textit{in situ} annealing procedures. The circular Fermi surface of the first light subband and two cigar-like Fermi surface sheets can be seen in all cases, just as found on the fractured STO (001) surface. The characteristic features of the electronic structure and in particular the presence of multiple orbitally-ordered subbands, do not change depending on the annealing procedure. This indicates a remarkable universality of the STO (001) 2DEG subband structure. It is insensitive to the bulk doping level of the single crystal, the origin of the residual bulk doping, the marked differences in macroscopic surface roughness between fractured and polished surfaces and the various terminations and reconstructions yielded by different annealing procedures.

\subsubsection{Bandstructure Modelling}
\label{models}

\begin{figure}

	\centering

			\includegraphics[width =0.75\textwidth]{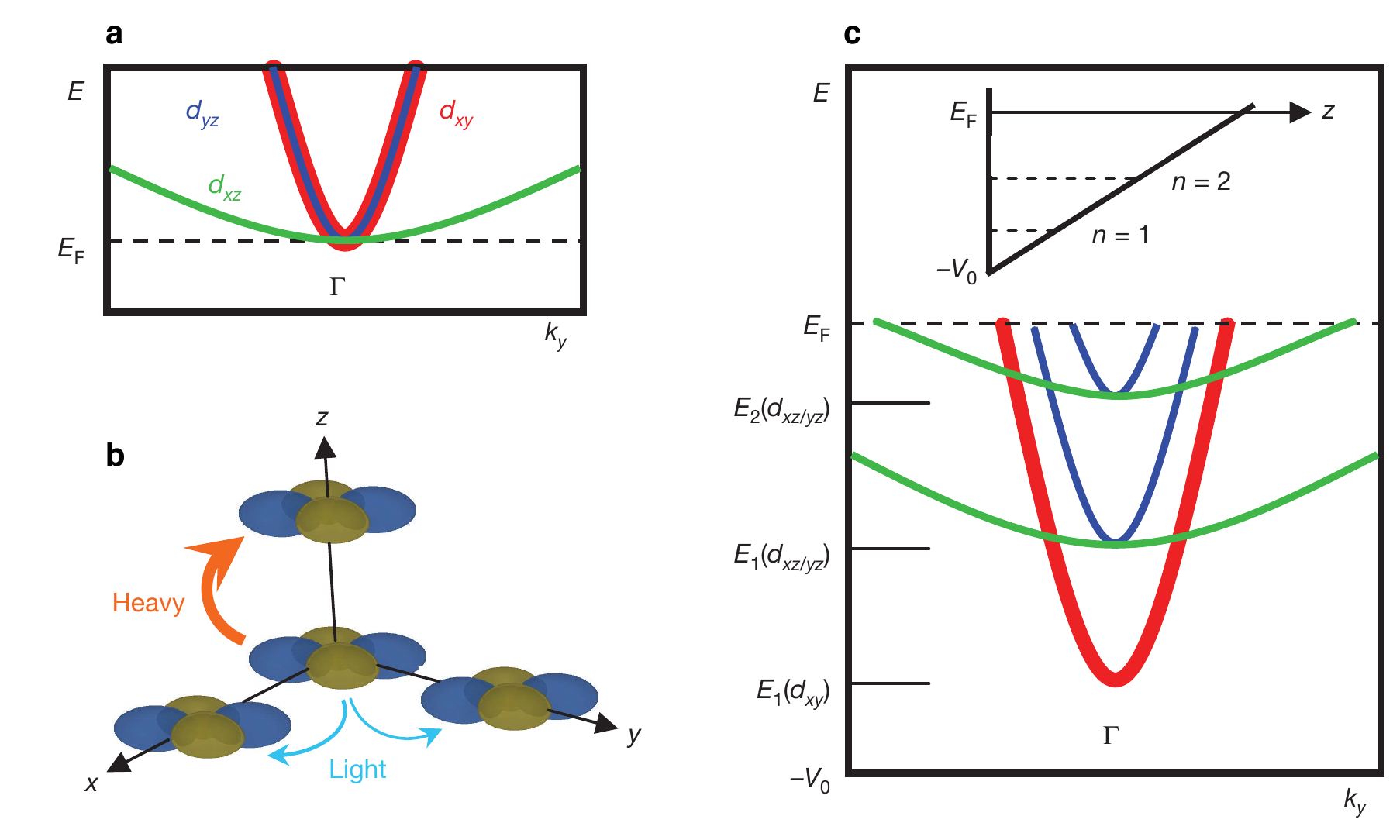}
	\caption{Modelling the effects of quantum confinement in a wedge-shaped potential well. (a) Simplified bulk band structure of STO along the $k_y$ axis. The colours indicate the orbital charater as indicated. (b) Cartoon of the allowed motion of an electron between \dxy{} orbitals in a cubic structure. For a band formed of such electrons, the direction in which the effective band mass would be "heavy" or "light" is indicated. This is determined by the small or large overlap of the orbital in that direction, respectively. (c) The subband structure expected using the approximate solution of a particle of mass $m_z$ in a wedge potential to define the subband confinement energies at the $\Gamma$ point. The inset shows the wedge potential profile as a function of $z$, the direction perpendicular to the crystal surface. From Santander-Syro \textit{et al.} \cite{Santander-Syro2011a}.}
	\label{SS2011_F1}
\end{figure}

Santander-Syro~\textit{et al.}~\cite{Santander-Syro2011a} first pointed out that the energetic ordering of the subband ladder in the STO (001) surface 2DEG can be understood qualitatively with a simple model of quantum confinement of strongly anisotropic bands with \tg{} orbital character. As illustrated in \figref{SS2011_F1}(b), each \tg{} orbital is associated with electron motion with a light effective mass along two crystallographic axes and a heavy mass along the third direction. It follows that 2DEG subbands with a heavy in-plane effective mass derive from orbitals with light out-of-plane mass $m_z$ while light subbands can have either a light or heavy out-of-plane mass. If bands with these orbital characteristics are subjected to a simple wedge potential as sketched in the inset of \figref{SS2011_F1}(c), they experience an energy shift proportional to $m_z^{-1/3}$. This shift relative to the bottom of the potential well will be smallest for the \dxy{} band which has light effective mass in the surface plane and a heavy mass along the confinement direction. Thus the lowest order \dxy{} band will sit near the bottom of the wedge potential. Conversely the bands with out-of-plane \dxzyz{} orbital character will be pushed to higher energy. This is shown in \figref{SS2011_F1}(c) and qualitatively reproduces experimental results from ARPES at the STO surface and from X-ray linear dichroism (XLD) of the LAO/STO interface \cite{Salluzzo2009}. It is clear however that this simple wedge model cannot accurately predict the relative subband energies seen in ARPES experiments.

\begin{figure}
	\centering

	\includegraphics[width =1\textwidth]{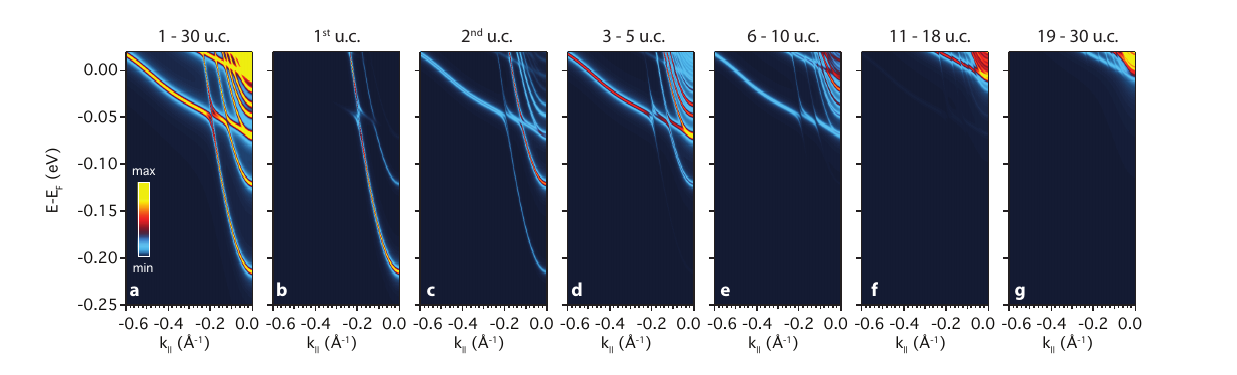}
	\caption{Layer-projected tight binding superecell calculations of the electronic structure of the STO (001) surface 2DEG integrated over (a) the full 30 unit cell supercell, and (b)-(e) individual or few unit cell regions as labelled in the figure. High intensity (yellow) corresponds to a high wavefuncion weight. This calculation is based on and \textit{ab initio} DFT parametrization of bulk STO, including a field dependent dielectric constant as proposed in references \cite{Stengel2011,Copie2009} and implements electronstatic boundary conditions that reproduce the overall bandwidth of the 2DEG. From King \textit{et al.} \cite{King2014}.}
	\label{King2014_SF3}
\end{figure}

More realistic confinement potentials can be obtained from a self-consistent solution of the Poisson and Schr\"odinger equations~\cite{Meevasana2011a,Stengel2011}. Using an accurate tight-binding parametrization of the bulk conduction band and including surface band bending as an on-site potential term in a large supercell extending several tens of unit cells perpendicular to the surface, these calculations reproduce experimental band structures in various quantum confined systems to a high degree of accuracy \cite{King2008,Bahramy2012a,King2012a,King2014,McKeown2016,Gariglio2015,Yukawa2015}.
The subband structure for such a calculation based on an \textit{ab initio} DFT calculation of bulk cubic STO including spin-orbit coupling, is shown in \figref{King2014_SF3}(a). The electrostatic boundary conditions are chosen such that the total calculated bandwidth matches the saturated bandwidth of the 2DEG. The resulting confinement energies of the subbands are in good agreement with ARPES data as demonstrated in \figref{rashba}(a). Such calculations can be used to estimate the finite extent of the 2DEG along the confinement direction by projecting the eigenstates onto atomic layers as shown in \figref{King2014_SF3}~\cite{King2014}. The confinement energy shifts are directly related to the spatial extent of the wave functions perpendicular to the surface, giving the intuitive result that the L1 \dxy{} band is very spatially confined, as indicated by the high intensity of the band in \figref{King2014_SF3}(b). The wavefunctions of higher order \dxy{} bands are peaked successively further below the surface and the high intensity of the H1 subband in \figref{King2014_SF3}(d) shows that the wavefunction of the first \dxzyz{} band is centred 3-5 unit cells below the surface. The distribution of intensity for the heavy band shows that it is also much more spatially extended than the \dxy{} bands. This subband-specific spatial profile of the wavefunctions reflects the form of the potential well and is consistent with the matrix element structure seen in \figref{Wang2016_SF2} and discussed in \secref{STO_001}. From this it is clear that the spatial extent of the 2DEG as a whole is not a single well defined quantity. 
While the total charge density will always peak near the surface/interface, it can have very long tails extending deep into the bulk. These tails are often not seen in spectroscopic experiments whose signal strengths are typically proportional to the charge density. However, they might have a strong influence on transport properties which are often dominated by the most mobile carriers that might reside far from the interface where scattering is generally lower. We also note that the shape of the confinement potential and  total carrier density distribution in oxide 2DEGs  is a strong function of carrier density. This is particularly true for STO since the strong suppression of its dielectric constant in an electric field progressively enhances the confinement at high carrier density. These considerations might explain why transport measurements of the LAO/STO (001) 2DEG often find thicknesses $>10$~nm~\cite{Basletic2008,Reyren2009}, while spectroscopic studies and DFT calculations typically report a confinement of the 2DEG within $<2$~nm for both surface and interface systems~\cite{Sing2009,Dubroka2010,Cancellieri2013b,King2014,Stengel2011}.
	
Full \textit{ab-initio} electronic structure calculations of STO surface 2DEGs have been presented in Refs.~\cite{Altmeyer2015,Garcia-Castro2015} and qualitatively agree with the results of tight-binding supercell calculations. While such calculations are suitable for studying the behaviour of oxygen vacancies, direct comparison of the resulting band structure with experiment is hindered by limitations in the size of the supercell and the ordered nature of vacancy arrangements in density functional calculations that impose translational invariance in the surface plane.

\subsubsection{Rashba Spin Orbit Coupling}

The confinement potential associated with a surface 2DEG inherently breaks inversion symmetry, which lifts the constraint of spin degeneracy from the band structure. Spin splitting may result due to the coupling of an electron's motion to its spin via the effective in-plane magnetic field resulting from a Lorentz transformation of the symmetry-breaking electric field at the surface. This is known as the Rashba spin-orbit interaction \cite{Rashba}. The magnitude of spin-splitting in a Rashba system is linearly proportional to the strength of the electric field. This behaviour is observed in conventional semiconductor 2DEGs \cite{nitta1997} and is a prerequisite for many applications in spintronics such as the spin-field effect transistors (spin-FET) \cite{Koo2009}. %permits electrostatic control of the spin degree of freedom thereby presenting interesting opportunities for the realisation of spin-field effect transistors (spin-FET) \cite{Koo2009}.
Rashba spin-splitting is also expected to scale with the strength of the atomic spin-orbit interaction (SOI) in the host material. Therefore considering the  light atomic masses of the constituent elements of STO and consequently small atomic SOI, a small Rashba effect might na\"ively be expected in STO based 2DEGs. Surprisingly though, a substantial gate-tunable spin splitting of $2-10$ meV has been deduced from transport experiments for both the LAO/STO interface 2DEG \cite{Caviglia2010,Hurand2015,Fete2012,Liang2015} and electrolyte-gated STO \cite{Nakamura2012}. The spin-splitting was found to have a strongly non-linear dependence on gate-field and weak antilocalization measurements suggest that it is proportional to $k^3$ rather than being $k$-linear as expected in the simplest models of Rashba spin splitting \cite{Nakamura2012}. 

\begin{figure}
	\centering

	\includegraphics[width=0.6\textwidth]{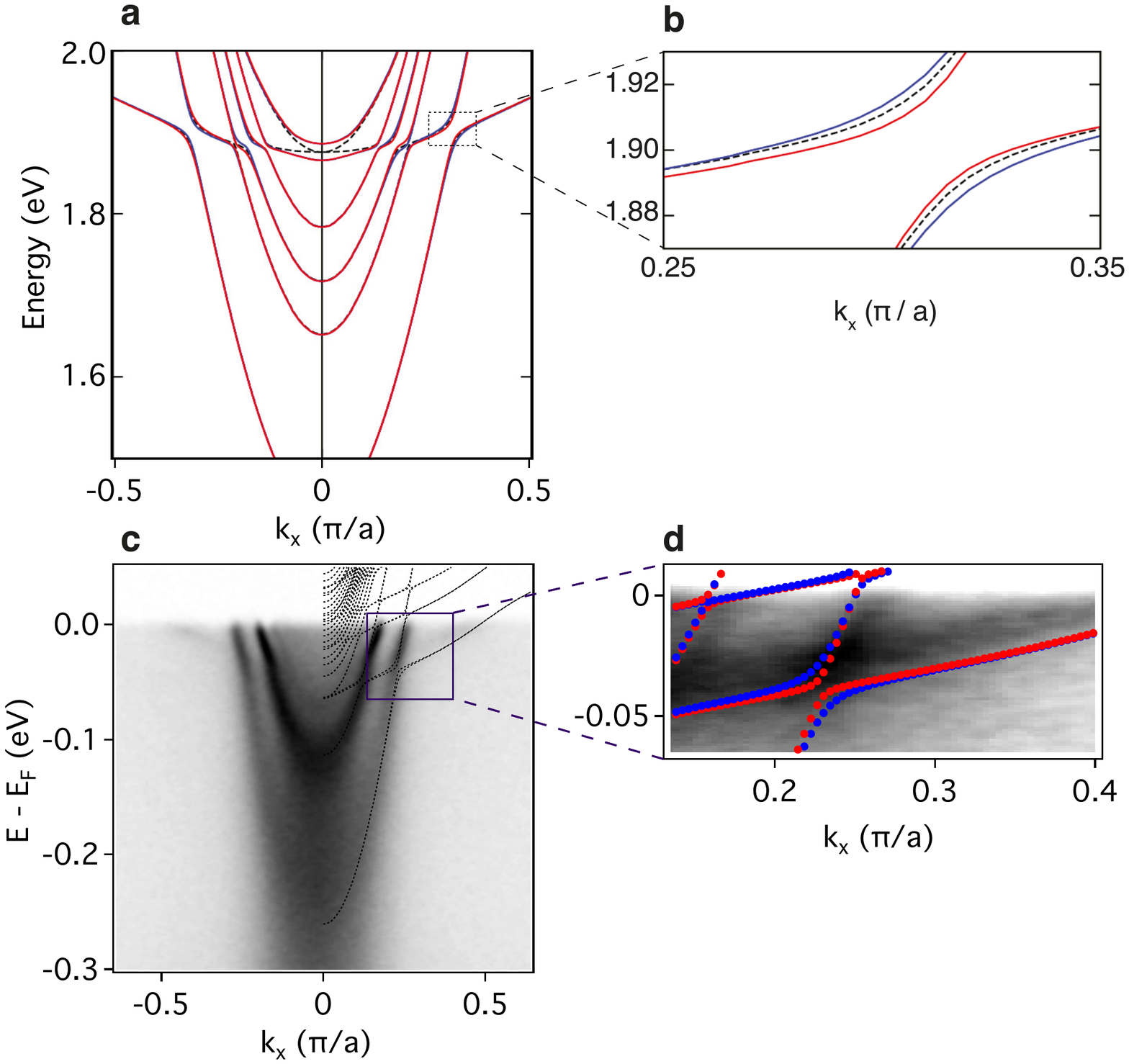}
	\caption{Unconventionl Rashba-like spin splitting of STO (001) 2DEG subband structures. (a) DFT band structure of a single $n$-type vacuum/LAO/STO interface. (b) The boxed region in (a) is enlarged which reveals that the bands are not spin degenerate, and that spin-splitting is enhanced at the avoided crossings of light and heavy subbands. (c) Band structure of the STO surface found from tight-binding supercell calculations based on \textit{ab initio} DFT electronic structure of bulk STO (black lines), overlayed on ARPES data (grascale image) of the STO (001) surface2DEG. This shows good agreement with the experimental confinemnt energies and many similarities with the interface calculation in (a). (d) The boxed region in (c) is enlarged and shows a similar spin-splitting as seen in (b) for the LAO/STO interface. In (a),(b) and (d) opposite in-plane spin channels are coloured red and blue. Adapted from Zhong \textit{et al.} \cite{Zhong2013b} and McKeown Walker \textit{et al.} \cite{McKeown2016}.}

	\label{rashba}
\end{figure}

Using band structure calculations based on relativistic DFT shown in \figref{rashba}(c) and (d), Zhong \textit{et al.} \cite{Zhong2013b} demonstrated that these behaviours can be understood as the signatures of an unconventional Rashba spin-splitting arising from the multi-orbital nature of 2DEGs in STO. \figref{rashba}(d) shows the band structure for a single LAO/STO interface. In the region of the avoided crossings of light and heavy subbands the spin splitting is dramatically enhanced and clearly deviates from a $k$-linear form. A strikingly similar spin structure of the STO surface 2DEG can be seen in the tight binding supercell (TBSC) calculations of references \cite{King2014} and \cite{McKeown2016} as shown in \figref{rashba}(a) and (b) and has been found by several other theoretical studies \cite{Khalsa2013a,King2014,Kim2014a,Kim2013a,Shanavas2016,VanHeeringen2013,Zhou2015}. The authors of reference \cite{King2014} related this enhancement to the finite orbital angular momentum that arises at the crossings of bands of different orbital character and augments the spin-orbit interaction at these particular locations in momentum-space. In this case the gate voltage used in transport experiments not only directly tunes the Rashba coefficient in STO 2DEGs, but indirectly tunes the spin splitting by controlling the band filling and band structure. Indeed this indirect electrostatic tuning may be more important in such a system since the spin splitting is enhanced by approximately an order of magnitude at the avoided crossings of the \dxzyz{} and \dxy{} bands. However, this unconventional spin-splitting never exceeds $\sim10$ meV and thus remains below the resolution of high-resolution ARPES measurements such those of \figref{McK2015_F1}.

 In order to gain direct spectroscopic insight into the spin structure of the STO (001) surface 2DEG Santander-Syro \textit{et al.} \cite{Santander-Syro2014} and McKeown Walker \textit{et al.} \cite{McKeown2016} performed spin and angle resolved photoemission spectroscopy (SARPES) experiments. However, these authors reported conflicting results. McKeown Walker \textit{et al.} measured no significant spin polarization of the photocurrent above the $\sim5\%$ noise level of their experiment. Considering the complex subband structure and poor experimental resolution in SARPES, they reason that such a negligible photocurrent polarization is fully consistent with the spin splitting shown in \figref{rashba}. However, direct experimental confirmation of this unconventional Rashba effect remains elusive. Conversely Santander-Syro \textit{et al.} measured a large polarization of the photocurrent which prompted them to propose an entirely new interpretation of the universal subband structure measured at the surface of STO (001) crystals. They propose that bands L1 and L2 are the fully spin polarized components of a single \dxy{} subband. In a Rashba system the spin states must be degenerate at the Brillouin zone centre. However, it is well established that L1 and L2 of the STO surface 2DEG are not degenerate. Santander-Syro  \textit{et al.} speculate that the presence of ferromagnetic domains, which generate a large Zeeman-like term lifting the $\Gamma$ point degeneracy, could account for this. Indeed some DFT calculations for ordered oxygen vacancy arrangements at the STO (001) surface show magnetic solutions of comparable energy to the paramagnetic case \cite{Altmeyer2015,Garcia-Castro2015}. However, in these calculations the remnant in-plane spin component due to the Rashba effect is an order of magnitude smaller than that measured by Santander-Syro \textit{et al.}. To date, the origin of the discrepancy between these two SARPES experiments is unclear and the details of the spin structure remain elusive.

\subsection{\sto{} (111) and (110) surface 2DEGs}
Two-dimensional electron gases in ABO$_3$ transition metal oxides oxides are by no means restricted to the (001) plane. 2DEGs have been successfully engineered at the bare (111) and (110) surfaces of \sto{} \cite{Wang2014a,McKeown2014,Rodel2014} and at interfaces with these orientations \cite{Herranz2012,Annadi2013,Raghavan2015}. These studies are motivated, in part, by theoretical predictions of novel ferromagnetic and ferroelectric states and topological phases in (111) bilayers of cubic perovskites \cite{Xiao2011,Doennig2013} and the intrinsic in-plane anisotropy of the (110) plane. In the following we will discuss the overall electronic structure of (111) and (110) orientated surface 2DEGs on STO and relate it to the framework for quantum confinement developed in \secref{models} and its interplay with the different symmetries of these surfaces. 

\begin{figure}
	\centering
			\includegraphics[width =0.9\textwidth]{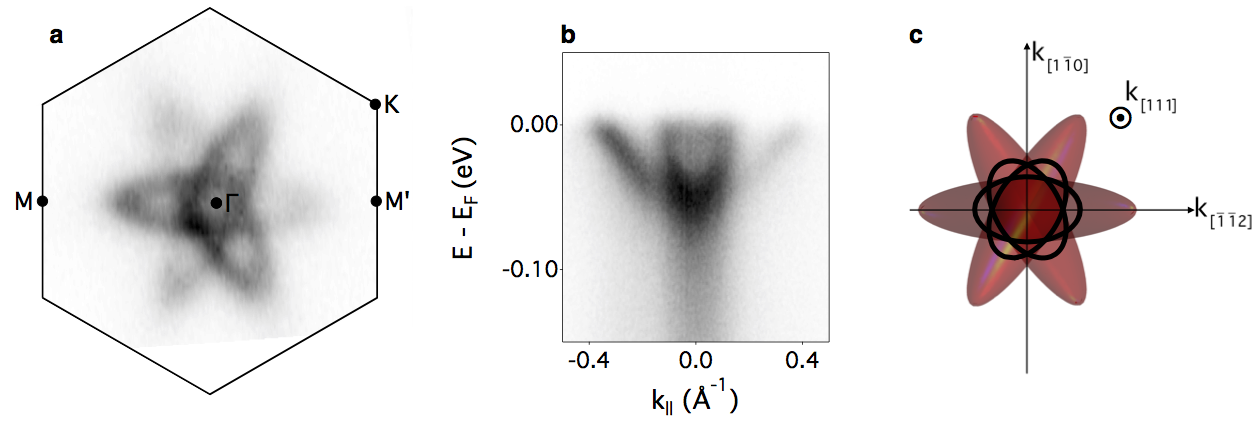}
	\caption{Electronic structure of the STO (111) surface 2DEG. (a) Fermi surface showing three elliptical Fermi surface sheets. The black hexagon indicates the surface Brilluion zone and the Black dot indicates the $\Gamma$ point. (b) Dispersion along the $\Gamma$-M$^\prime$ direction. Spectral weight at the Fermi level is from a second subband. (c) Projection onto the (111) plane (red shape) and cut at the $\Gamma$ point (black lines) of a model STO bulk Fermi surface. The form of the STO (111) 2DEG Fermi surface closely ressembles the projection. Adapted from McKeown Walker \textit{et al.} \cite{McKeown2014}.}
	\label{McK2014}    
\end{figure}
\begin{figure}
	\centering
	\includegraphics[width =0.8\textwidth]{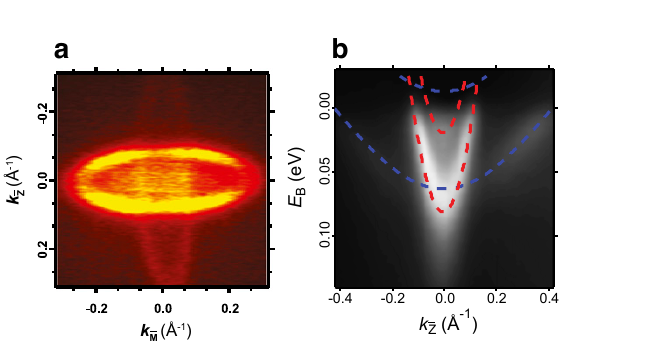}
	\caption{Electroninc structure of the STO (110) surface 2DEG. (a) Fermi surface showing two intersecting elliptical Fermi surface sheets. The \dxzyz{} band are degenerate and have higher intensity here due to the polarization of the exciting radiation. (b) Disperion along the $k_{\bar Z}$ direction. The dashed red(blue) lines indicate the dispersion of the \dxzyz(\dxy) states and possible higher order subbands. Adapted from Wang \textit{et al.} \cite{Wang2014a}.}
	\label{Wang2014_F3F4}  
\end{figure}

\figref{McK2014} (a) shows the Fermi surface of the STO (111) surface. Three intersecting elliptical Fermi surface sheets with an overall six-fold symmetry can be seen. Each of these can be associated with the projection of a single \tg{} component of the conduction band onto the (111) plane. As seen in \figref{McK2014} (b), which shows the bands dispersing along the long axis of one ellipse, within the accuracy of the experiment these electron like bands are degenerate at the $\Gamma$ point. This is a natural consequence of the $120^\circ$ rotational equivalence of the \tg{} orbitals in the (111) plane, which causes all three bands to have the same effective mass along the confinement direction. When applied to the STO (110) plane, these arguments predict degenerate bands of \dxzyz{} character, due to the "semi-heavy" hopping for these orbitals, and a \dxy{} band with weaker confinement. This is indeed the case, as seen in \figref{Wang2014_F3F4}(a) and (b) which shows the Fermi surface and a band dispersion from the STO (110) 2DEG \cite{Wang2014a}. Thus the STO (110) 2DEG is orbitally polarized, although the relatively small variation between the band masses for the \dxzyz{} and \dxy{} bands along the [110] direction leads to a much less dramatic orbital reconstruction than found for the STO (001) 2DEG.

Notably the two dimensional carrier densities of the fully saturated 2DEGs on all three low index surfaces of STO are comparable to each other, with $n_{2D}\sim2\cdot10^{14}$~cm$^{-2}$. Additionally, models reproducing the saturated band structures of both the (111) and (001) 2DEGs find very similar confinement fields at the surface \cite{McKeown2014,King2014}. This suggests a common origin for the saturation of the 2DEG bandwidth controlled by a physical limit on the electric field strength at the surface. One possible mechanism for such a limit could be the onset of diffusion of charged oxygen vacancies as the electric field increases. The common carrier density implies that 2DEGs on different low index surfaces will have different bandwidths and spatial extents. This is evident from the dispersion plots of Figs. 11,10 and 4 and photon energy dependent measurements that demonstrate the extended nature of the electron density of (110) and (111) 2DEGs, in analogy to the heavy \dxzyz band of the STO (001) 2DEG.

 Both the STO (111) and (110) surfaces show dramatic in plane subband mass enhancements with respect to the dispersion of bulk carriers along the same direction.  As shown in Refs. \cite{McKeown2014,Wang2014a} this is an intriguing effect of quantum confinement, rather than a manifestation of many-body interactions. Intuitively this mass enhancement can be understood as the result of the projection of the three dimensional bulk Fermi surface onto a 2D plane. As illustrated in \figref{McK2014} for the (111) plane, the contours of the projection (red) are elongated with respect to those of the bulk dispersion in that plane (black lines) and thus correspond to higher effective masses. An alternative way of thinking is that the in-plane subband mass is enhanced as a result of the zig-zag hopping of carriers moving in the surface plane, which, in conjunction with surface band banding leads to reduced hopping elements and thus enhanced effective masses in-plane.

  Along both the [111] and [110] directions \sto{} can be viewed as a stack of charged planes. Therefore the charge accumulation mechanism at (111) or (110) surfaces might be expected to be different from what is observed for the neutrally stacked [001] direction. For example, surface reconstructions that compensate the polarity of the surface could induce spontaneous charge accumulation. However the origin of the (111) and (110) surface 2DEGs has very much the same phenomenology as the (001) surface, as discussed in \secref{OV_section}. Additionally, there is no evidence from ARPES that surface reconstructions influence the (111) and (110) surface 2DEG band structures. For the case of the STO (110) 2DEG where the surface is known to have a $4\times1$ reconstruction, this has been attributed to the protective nature of the insulating over-layer of titania formed by this reconstruction. Wang \textit{et al.}  proposed that light-induced oxygen vacancies migrate below the overlayer and dope electrons which are not perturbed by the potential of the surface reconstruction. For the case of the STO (111) 2DEG, where details of the surface termination are not known, the insensitivity of the states to possible reconstructions was attributed to the wavefunctions of the 2DEG being centred far below the surface. The envelope wavefunction solutions of the self-consistent tight binding supercell calculations \cite{McKeown2014} show the wavefunctions peak $\approx6$ Ti layers below the surface. This is in line with the observation that back-folded bands have been observed for the 2DEGs at the reconstructed STO (001) and anatase \tio{} (001) surfaces, where the \dxy{} subbands are more tightly confined at the surface \cite{Wang2016a}. It may also provide these systems with some degree of insensitivity to surface or interface impurities.

%

% not like dxy in 001 where they "like to be confined because they are more or less already 2D
%wavefunction peaked 6 Ti layers below surface

%periodicity is that of a single Ti layer

\section{Surface 2DEGs in other transition metal oxides}
The diverse bulk properties of the large number of transition metal oxides suitable for heteroepitaxy hold much potential for both fundamental studies and applications. Inducing 2DEGs in insulating oxides other than STO is an important step towards unlocking this potential. However, so far  little is known about the prerequisites on host materials and interface properties required to this end. To date, besides on \sto, highly itinerant surface 2DEGs have been observed on \kto~\cite{King2012a,Santander-Syro2012,Bareille2014b} and anatase \tio~\cite{Rodel2015,Wang2016a} which both have an empty $d$-shell in stoichiometric form, while attempts to induce a 2DEG in rutile \tio{} and the Mott insulator \lto{} were unsuccessful~\cite{Rodel2015,McKeownThesis}. In the case of rutile \tio, this might be related to the strong tendency of bulk samples to localize excess carriers. Hole doped bulk \lto{} on the other hand, is known to host itinerant carriers~\cite{Tokura1993} suggesting that its fundamental material properties should not prohibit the formation of 2DEGs. This highlights one of the key challenges of this field. While it is clear that the creation of a 2DEG requires chemical doping or charge transfer across an interface, it is often hard to predict whether carrier doping of insulating oxides induces metallicity.

\subsection{\kto}
\label{kto}
\kto{} shares important properties with \sto. Both are \abo{} perovskites with empty $d$-shell and are close to a ferroelectric instability. However, unlike in STO, the bulk truncated (001) surface of KTO is strongly polar. Additionally, in KTO the effective masses of the conduction band are lighter and the spin-orbit interation in the Ta 5$d$ shell is more than an order of magnitude larger than in the Ti 3$d$ states. This suggests that KTO might be a suitable material on which to engineer 2DEGs with high mobility and large tunable Rashba splitting. However, the lack of established surface preparation recipes resulting in well-ordered surfaces with single termination and the poor stability of the KTO (001) surface at high temperature have thus far prohibited the growth of heteroepitaxial interfaces with a quality as it is routinely achieved with STO substrates. Indeed, KTO based 2DEGs were first induced with a parylene gate dielectric~\cite{Nakamura2009} and by electrolyte gating~\cite{Ueno2011}, while the first oxide interface inducing a 2DEG in KTO was only reported recently~\cite{Zou2015}. Notably, these studies found superconductivity at high carrier density~\cite{Ueno2011} and reported spin-precession lengths that are significantly shorter than in InGaAs and tunable over a very wide range varying from 20 to 60~nm with gate voltage~\cite{Nakamura2009} suggesting much potential of KTO for spintronic devices. Yet, the Rashba effect which causes the spin-precession and even the overall band structure of KTO based 2DEGs remain poorly understood.

  King \textit{et al.} reported a 2DEG on the bare (001) surface of KTO and studied its band structure with a combination of ARPES and tight-binding supercell calculations~\cite{King2012a}. The experimental data showed an occupied band width of $\sim400$~meV and resolved two isotropic light subbands with effective masses of $\sim0.3$~m$_{e}$ and a shallower subband with $m^{*}\sim2-3$~m$_{e}$ contributing an elliptical Fermi surface. These results were confirmed by Santander-Syro \textit{et al.}~\cite{Santander-Syro2012} and could largely be reproduced by band structure calculations although the agreement is not as good as in STO (001). In particular, finer details, such as hybridization gaps between the subbands and the theoretically predicted Rashba splitting could not be resolved experimentally. From the line width, King~\textit{et al.} deduced a upper limit for the Rashba splitting of $\sim0.02$~\AA$^{-1}$~\cite{King2012a}, which is consistent with the spin precession lengths reported in Ref.~\cite{Nakamura2009} but still more than an order of magnitude lower than in other surface systems containing heavy atoms, such as the L-gap surface state on Au(111) \cite{LaShell1996} or the surface 2DEG on the topological insulator Bi$_2$Se$_3$~\cite{King2011}.

The authors of Ref.~\cite{King2012a} attributed the modest Rashba splitting in KTO to the particular orbital character of the bulk states from which the 2DEG derives. The strong spin-orbit interaction in KTO restores the orbital angular momentum, which is largely quenched in the 3$d$ counterpart STO. Its conduction band is thus more appropriately described by total angular momentum states rather than the crystal field eigenstates commonly used in STO. The conduction band edge at the $\Gamma$ point of KTO is formed by a quartet of $J_\mathrm{eff}=3/2$ states, which are a linear combination of all three \tg{} orbitals. The $J_\mathrm{eff}=1/2$ doublet is split off by the spin-orbit gap $\Delta_{\mathrm{SO}}\approx 400$~meV. This splitting is larger than the occupied band width of KTO based 2DEGs reported in the literature~\cite{King2012a,Santander-Syro2012}, which therefore derive from the $J_\mathrm{eff}=3/2$ states. To a first approximation, the 2DEG in KTO has thus the same orbital composition as 2D hole gases in typical III-V semiconductors such as GaAs which suppresses the k-linear Rashba term~\cite{Winkler2003}. It should be noted however that, due to the same physics that produces orbital ordering in the STO (001) 2DEG, the strong quantum confinement in KTO (001) modifies the orbital character and re-introduces a significant orbital polarization as pointed out in Ref.~\cite{Santander-Syro2012}.

\subsection{Anatase \tio}
\label{tio2}
\tio{} crystallizes in the rutile and anatase structures and is one of the most intensely studied TMOs due to its diverse applications in heterogeneous catalysis, photo-catalysis, gas-sensing or photovoltaics and its use as transparent conductive coating or simply as biocompatible white pigment. While the surface science of \tio{} is intensely studied~\cite{Diebold2003}, \tio{} has so far received less attention as a host material for oxide 2DEGs and only a few studies reported conductive interfaces with other TMOs~\cite{Minohara2014,Sarkar2015}. Similar to STO, \tio{} is susceptible to the creation of light-induced oxygen vacancies~\cite{Knotek1978,Moser2013a}, which was exploited by Moser~\textit{et al.}~\cite{Moser2013a} to dope anatase bulk single crystals and PLD-grown thin films inducing quasi-3D electronic states which showed strong signatures of electron-phonon interaction in the ARPES spectra. Subsequently, R\"odel~\textit{et al.} showed that under appropriate conditions fully 2D quantum confined states with the characteristic subband ladder of a 2DEG can be induced on the (001) and (101) surface of anatase \tio~\cite{Rodel2015}. Intriguingly though, the same approach did not induce itinerant carriers in rutile \tio~\cite{Rodel2015}. 

\begin{figure}
	\centering
	\includegraphics[width =0.7\textwidth]{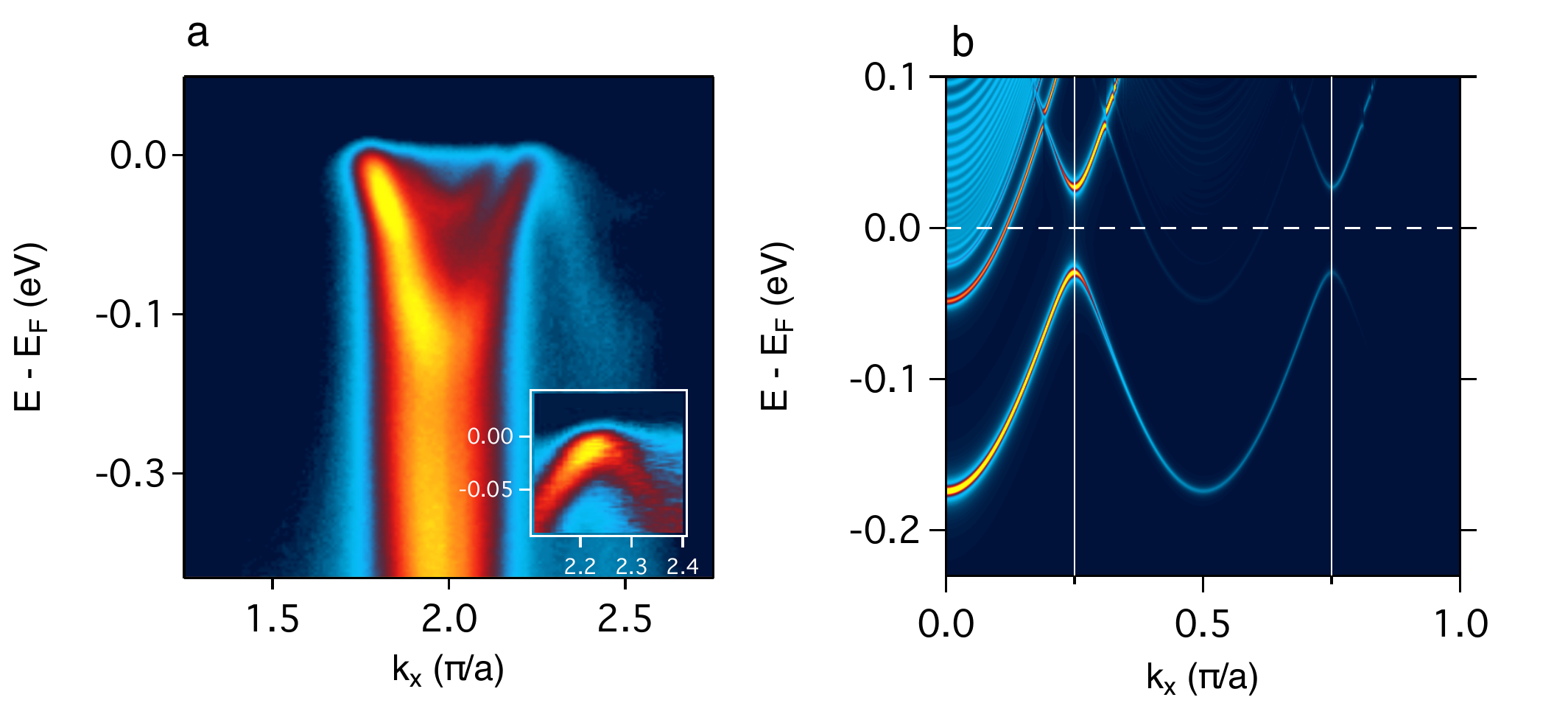}
	\caption{(a) 2DEG subband dispersion at the surface of anatase \tio{} (001) thin film terminated by a $(1\times4)$ surface reconstruction. Backfolding of the $n=1$ quantum well state at the superlattice zone boundary is highlighted in the inset where the opening of a superlattice band gap is observed. (b) Tight binding supercell calculation of the spectral weight distribution for a $4\times30$ supercell with an in-plane potential modulation of $\sim 100$~meV estimated from \textit{ab-initio} calculations of core-levels imposed in the topmost unit cell. For details see Wang \textit{et al.} \cite{Wang2016a}.}
	\label{Wang2016_TiO2}    
\end{figure}

Unlike in STO, the conduction band minimum of anatase \tio{} is of pure $d_{xy}$ orbital character with the other \tg{} orbitals split off by $\sim0.5$~eV. This results in a particularly simple electronic structure of anatase \tio{} based 2DEGs with isotropic Fermi surfaces of all subbands on the (001) surface and concentric elliptical Fermi surfaces on the (101) surface~\cite{Rodel2015}. The  $d_{xy}$ orbital character of anatase \tio{} based 2DEGs also leads to a particularly strong confinement of the carriers in a narrow layer below the surface. This was exploited by Wang~\textit{et al.} to demonstrate a periodic lateral modulation of the 2DEG by the $(1\times4)$ surface reconstruction of \textit{in-situ} grown anatase \tio{} thin films with (001) orientation~\cite{Wang2016a}. Tuning the Fermi wave vector of the first subband to coincide with the superlattice Brillouin zone boundary corresponding to the surface reconstruction, the authors of Ref.~\cite{Wang2016a} found a sizeable superlattice band gap at the Fermi level, as shown in Fig.~\ref{Wang2016_TiO2}. This suggests a new route towards electronic structure engineering in oxide 2DEGs by exploiting the ubiquitous surface reconstructions of TMOs.

%LAO/TiO2 (001) shows 2DEG for both rutile and anatase phase but in the rutile 2DEG carriers localize below 25 K \cite{Sarkar2015}

%\begin{figure}
%	\centering
%	\includegraphics[width =0.6\textwidth]{Figures/Rodel2015_F2.png}
%	\caption{anatase single crystal \tio{} (001) data from \cite{Rodel2015}}
%	\label{Rodel2015_F2}     % Give a unique label
%\end{figure}
\section{Many-Body Interactions in TMO 2DEGs}
\label{manybody}

The thermodynamic and transport properties of transition metal oxides are often dominated by many-body interactions. Prominent examples include high temperature superconductivity in cuprates, colossal magnetoresistance in manganites or the ubiquitous metal-insulator transitions in ultrathin TMO films~\cite{Yoshimatsu2010,King2014a}. Unlocking the full potential of TMO 2DEGs will require an improved understanding of these interactions as the interfacial carrier density is tuned, which is a formidable task. Here, we briefly review the first microscopic measurements of many-body interactions in oxide surface 2DEGs using ARPES~\cite{Moser2013a,King2014,Chen2015,Wang2016}. These studies all focus on electron-phonon interaction (EPI) in anatase \tio{} and \sto{} as the density of itinerant carriers is tuned by controlling the oxygen vacancy concentration using the methods described in \secref{OV_section}.

Electron-phonon interaction in STO dominates the mobility of interface 2DEGs at elevated temperatures~\cite{Mikheev2015}, contributes to the large thermoelectric coefficient of depleted 2DEGs~\cite{Pallecchi2015} and has been invoked as the pairing glue for superconductivity~\cite{Edge2015,Gorkov2015}. Yet, until recently little was known about its nature and strength. Due to its strongly ionic character, lightly doped STO was often considered a model system for the nonlocal Fr\"ohlich interaction describing the dielectric screening of an excess charge by longitudinal optical (LO) phonons~\cite{Devreese2009}. In this model, EPI is strongly peaked at small momentum transfer $\mathbf{Q}$ and dominated by coupling to the highest LO branch \cite{Devreese2009}. On the other hand, STO is well known for its large static dielectric constant implying soft transverse modes, which can eventually condense into a ferroelectric state~\cite{Itoh2000,Rowley2014}. A recent theoretical study found that coupling to such a soft mode near a quantum critical point reproduces the supercondcuting dome of STO~\cite{Edge2015}.

Early experimental studies of EPI in STO focused on doped bulk samples. Van Mechelen \textit{et al.}~\cite{VanMechelen2008} and Devreese~\textit{et al.}~\cite{Devreese2010} showed that a pronounced mid-infrared peak in optical spectra can be reproduced quantitatively by the Fr\"ohlich model with a moderate coupling constant of $\alpha\approx2$ deduced from the independently determined static and high-frequency dielectric constants respectively. In this picture, the system remains fully itinerant despite the relatively large mass enhancement from EPI of $m^{*}/m_\mathrm{band}\sim2-3$ and forms a liquid of large polarons. On the other hand ARPES, which gives more direct insight into EPI, provided conflicting results. Chang \textit{et al.}~\cite{Chang2010} reported signatures consistent with coupling to the highest LO phonon branch with frequency $\Omega_{LO,4}\approx 100$~meV, as observed by van Mechelen \textit{et al.}~\cite{VanMechelen2008}, while Meevasana \textit{et al.}~\cite{Meevasana2010} reported a perturbative EPI with much stronger coupling to a soft LO mode than expected in the Fr\"ohlich model. While this discrepancy was never fully resolved, we speculate that it arises at least partially from accidental surface doping in the latter study. 
The ARPES study by Moser \textit{et al.} of quasi-3D carriers in anatase \tio{} created by photo-induced oxygen vacancies reported replica bands characteristic of Fr\"ohlich polarons at low carrier density and a progressive screening of EPI with increasing density~\cite{Moser2013a} which has similarities to what is observed in STO (001) as will be described in detail in this section.

\begin{figure}
	\centering
	\includegraphics[width =1\textwidth]{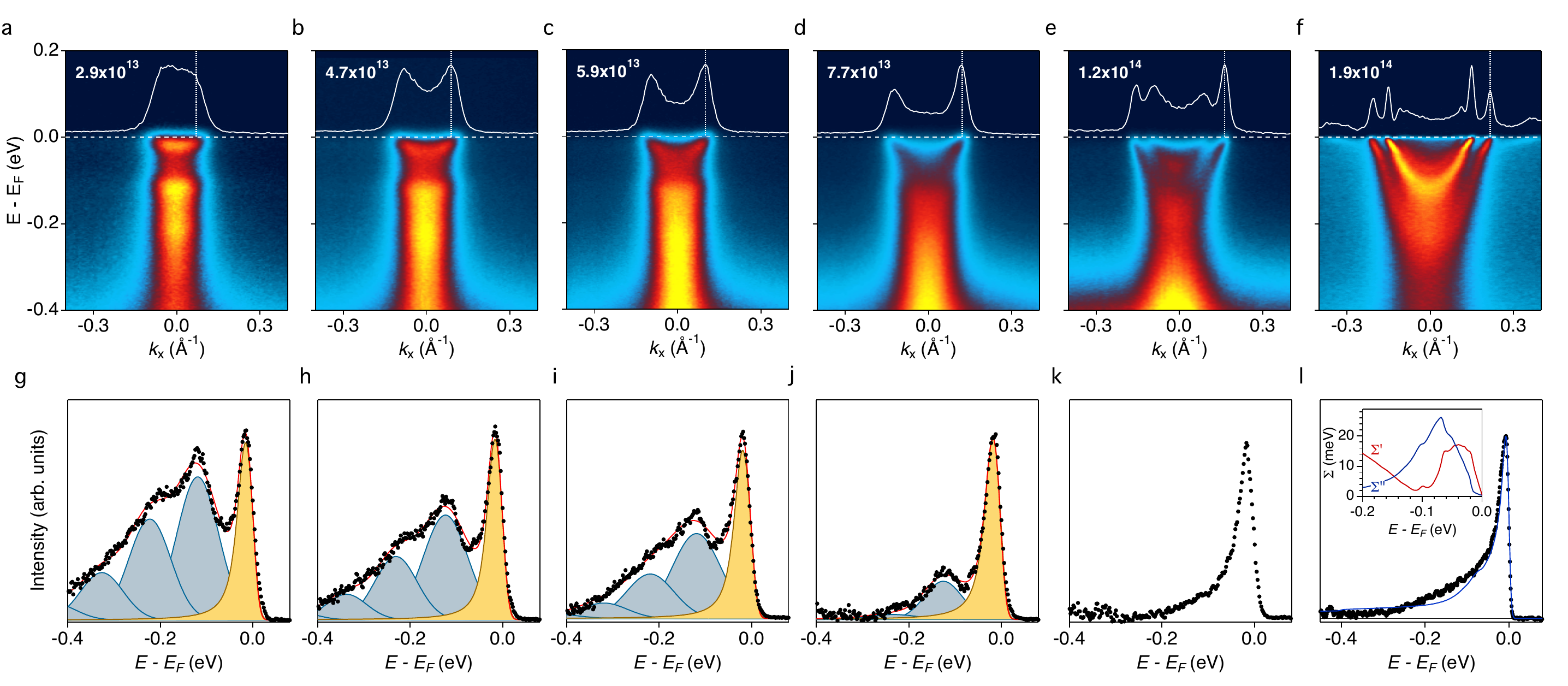}
	\caption{Evolution of the spectral function in the STO (001) surface 2DEG with increasing carrier density. (a-f) Raw dispersion plots and MDCs extracted at the Fermi level. Carrier densities are indicated in units of cm$^{-2}$. (g-l) EDCs at the Fermi wave vector $k_F$ indicated in the image plots. Fits in (g-j) use a Franck-Condon model with a single phonon mode of 100~meV. The thin blue line in (l) is a spectral function calculated for the conventional Eliashberg self-energy in the high-density limit reported in Ref.~\cite{King2014} and shown in the inset. Adapted from Wang \textit{et al.} \cite{Wang2016}.}
	\label{Wang2016_F2}   
\end{figure}

The marked influence of the carrier concentration on the spectral function is evident from the data on the STO (001) surface 2DEG of Wang~\textit{et al.}~\cite{Wang2016} reproduced in Fig.~\ref{Wang2016_F2}. Using \textit{in-situ} prepared surfaces, these authors could reduce the vacancy formation rate permitting a systematic study of the spectral function for carrier densities spanning nearly an order of magnitude from $\sim 3\cdot10^{13}\: -\: 2\cdot10^{14}$~cm$^{-2}$. At the lowest densities corresponding to an occupied quasiparticle bandwidth of $\sim20$~meV, the spectra of Ref.~\cite{Wang2016} show dispersive replica bands shifted by multiples of 100~meV to higher energy. This implies preferential coupling with small momentum transfer ($\mathbf{Q}\ll\pi/a$) to the LO$_4$ mode of STO, which is the hallmark of Fr\"ohlich polarons, quasiparticles formed by an excess electron dressed by a polarization cloud extending over several lattice sites that follows the charge as it propagates through the crystal~\cite{Devreese2009,Moser2013a,Chen2015}. Fig.~\ref{Wang2016_F2}(g) shows that the spectral weight at this density is clearly dominated by excitations involving one or more phonons (blue peaks in Fig.~\ref{Wang2016_F2}). The quasiparticle residue, which gives the relative spectral weight of the coherent quasiparticle (yellow in Fig.~\ref{Wang2016_F2}), was observed to be $Z\approx0.2$ corresponding to a coupling constant $\alpha\approx2.8$ in the Fr\"ohlich model~\cite{Mishchenko2000,Wang2016}, in fair agreement with the analysis of optical spectra in lightly doped bulk STO~\cite{VanMechelen2008,Devreese2010}. Comparing the Luttinger volume of the surface 2DEG with sheet carrier densities at the LAO/STO interface deduced from the Hall coefficient, the authors of Ref.~\cite{Wang2016} further concluded that interface superconductivity likely derives from a liquid of large polarons or possibly bipolarons.

As the density is increased, spectral weight is gradually transfered from the replicas to the quasiparticle band implying a decreasing effective coupling constant $\alpha$. Concomitant, the mass enhancement $m^{*}/m_{\mathrm{band}}$ decreases from 2.4~$m_e$ to 1.7~$m_e$, following the trend $m^{*}/m_\mathrm{band}=1/(1-\alpha/6)$ 
expected for Fr\"ohlich polarons at intermediate coupling strengths~\cite{Devreese2009}. 
%Here, $m_\mathrm{band}$ is the bare band mass of the light $xy$ sheet of $\sim0.6$~$m_e$. 
Using exact diagonalization, the authors of Ref.~\cite{Wang2016} showed that the observed evolution of EPI can be traced back to a gradual transition from dielectric screening at low density to dominantly electronic screening at high density qualitatively consistent with earlier ARPES measurements on quasi-3D states in oxygen deficient anatase \tio~\cite{Moser2013a}. Superconducting susceptibilities calculated within the same approach further showed that the dominant pairing channel has $s$-wave symmetry and indicated that a competition between the opposite trends of density of states and effective coupling strength underlies the dome shaped superconductivity observed at the LAO/STO interface~\cite{Caviglia2008}. 

It is worth noting that the effect of increasing carrier density in the STO (001) 2DEG is not limited to progressive screening of the long-range Fr\"ohlich interaction. At the saturation density of the surface 2DEG, the LO$_4$ mode is almost completely screened as is evident from the absence of any spectral signatures at 100~meV in Fig.~\ref{Wang2016_F2}(f). However, the coupling to lower frequency modes increases far beyond the predictions of the Fr\"ohlich model~\cite{Verbist1992,Meevasana2010,King2014} pointing at a remarkable complexity of EPI in oxide 2DEGs. We also point out that a significant coupling to the soft ferroelectric mode predicted in Ref.~\cite{Edge2015} cannot be excluded from published ARPES data on surface or interface 2DEGs in STO due to the difficulty of quantifying the quasiparticle dispersion at very low energy.

Electron-phonon interaction at the LAO/STO interface has recently been studied by tunneling spectroscopy and soft X-ray ARPES. Investigating tunnel junctions to interface 2DEGs with $\sim30$~meV occupied bandwidth, Boschker~\textit{et al.}~\cite{Boschker2015} found inelastic tunneling attributed to EPI with dominant coupling to the LO$_4$ mode and progressively weaker contributions from the softer LO modes of STO. Varying the chemical potential over $\sim5$~meV using a back gate did not change the coupling strength significantly.  While tunneling does not give absolute coupling strengths as they can be deduced from ARPES specta, the dominant contribution from the LO$_4$ mode is in agreement with the ARPES results on the STO (001) surface 2DEG for similar bandwidths~\cite{Wang2016}. The results on the surface 2DEG of Ref.~\cite{Wang2014a} were further confirmed by a soft X-ray ARPES study of an LAO/STO interface with $n_{2D}\approx 8\cdot10^{13}$~cm$^{-2}$~\cite{Cancellieri2015}, reporting a replica band and $Z\approx0.4$ in excellent agreement with the value reported in Ref.~\cite{Wang2014a} for the same density. This strongly suggests that the nature and strength of EPI is similar for interface and surface 2DEGs of the same density. We note, however, that manifestations of EPI in transport properties might differ between these two systems due to the different nature and density of defects.

%Dirk's isotope effect

%\begin{figure}
%	\centering
%	\includegraphics[width =0.5\textwidth]{Figures/Moser2013_F3.pdf}
%	\caption{Temperature dependence of angle-integrated photoemission spectra from oxygen deficient anatase \tio{} with a 3D carrier density of $n_{3D}\sim 5\cdot10^{18}$~cm$^{-3}$ showing clear replica bands at low temperature. The progressive broadening of the line width $\Delta E$ from the base temperature of 20~K up to 300~K is extracted in (b) and fitted with a Bloch-Gr\"uneisen form indicating a mass enhancement of $\lambda=\Delta E/2\pi k_{B}T\approx1.7$ (from Ref.~\cite{Moser2013a}).}
%	\label{Moser_F3}     % Give a unique label
%\end{figure}

\section{Discussion}
The study of 2DEGs at the surface of 3D transition metal oxides is clearly motivated by the importance of interface 2DEGs for oxide electronics. However, since the first discovery of surface 2DEGs on STO (001) in 2011~\cite{Meevasana2011a,Santander-Syro2011a}, their investigation by ARPES has, to some extent, evolved into its own sub-field with a number of interesting results from several groups as summarized in this chapter. These  include the comprehensive characterization of the subband masses, the subband ordering and the resulting orbital polarization for different surface planes and the rationalization of these effects in terms of quantum confinement~\cite{Meevasana2011a,Santander-Syro2011a,King2012a,King2014,McKeown2014,Wang2014a}; the identification of light-induced oxygen vacancies as the microscopic origin of the 2DEG carriers which has permitted tuning of the carrier density over a wide range~\cite{Meevasana2011a,McKeown2014,McKeown2015}; and the observation of a complex evolution of electron-phonon interaction with carrier density giving rare microscopic insight into the many-body interactions governing important properties of oxide 2DEGs~\cite{King2014,Wang2016,Chen2015}. Most of these studies have focused on different surface orientations of STO but 2DEGs have also been induced and studied at low-index surfaces of KTaO$_3$ and anatase TiO$_2$~\cite{King2012a,Santander-Syro2012,Rodel2015,Wang2016a}.

\begin{figure} 	
 	\centering
 	\includegraphics[width =0.9\textwidth]{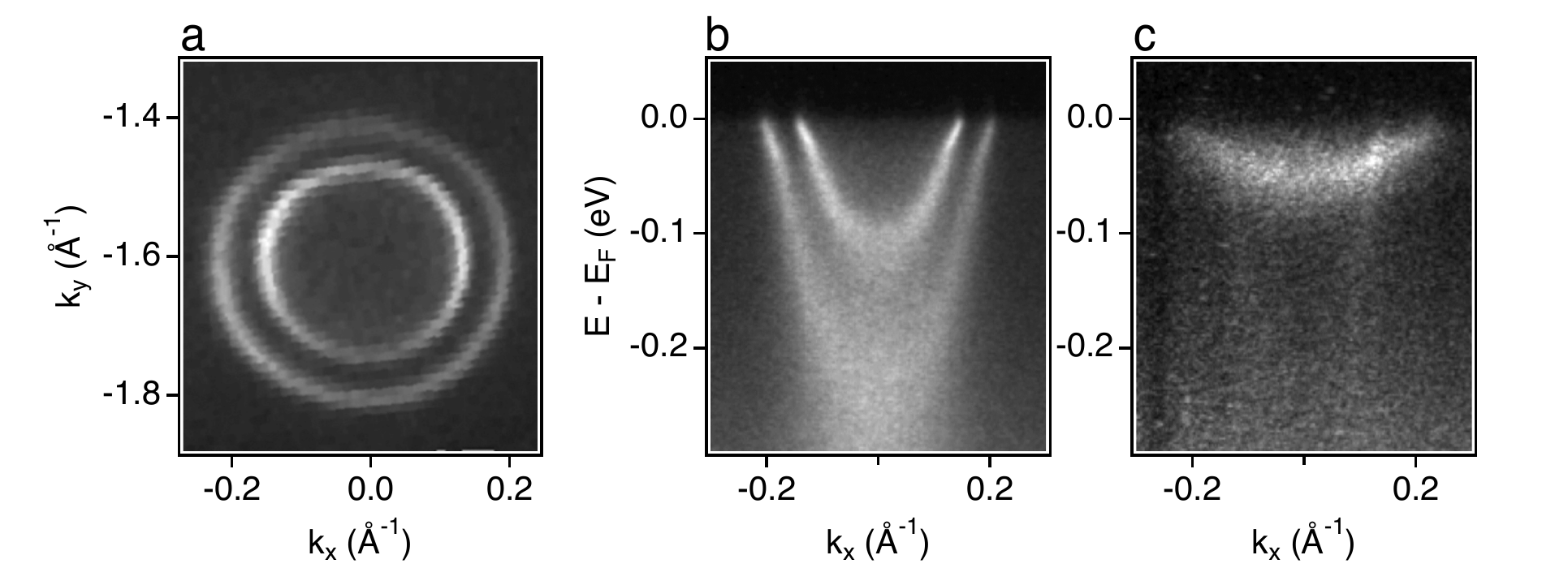}		
 	\caption{Electronic structure of the 2DEG at the interface of STO (001) and aluminium oxide formed when 2\AA{} Al is deposited on the bare \tio{} terminated (001) surface of  STO at room temperature. A redox reaction at the interface occurs and the surface of STO becomes reduced and doped with itinerant electrons which have the Fermi surface shown in (a). The Luttinger volume of these two concentric circular Fermi surface sheets is as found for the 2DEG at the bare (001) surface. The electron dispersions along the $k_x$ axis are shown in (b) and (c) for orthogonal polarizations of the exciting radiation revealing the light \dxy{} and heavy \dxzyz{} bands respectively. This orbital polarization is the same as found for the 2DEG at the bare (001) surface. Adapted from R\"odel \textit{et al.} \cite{Rodel2016}.}
 	\label{Rodel2016_F1}     
 \end{figure}

While these results are interesting in their own right, their relation with the properties of interface 2DEGs is not always clear. This being said, for the intensely studied STO (001) 2DEG a number of experiments suggest that some important properties are universal in the sense that they are largely determined by the host material and crystallographic orientation, rather than by structural details or the origin of the charges. For instance, orbital polarization is evident in the surface 2DEG~\cite{Santander-Syro2011a,King2012a,King2014} and has also been measured directly for the LAO/STO interface using X-ray linear dichroism~\cite{Salluzzo2009}. The Rashba spin-splitting of the surface 2DEG discussed in Refs.~\cite{King2014,McKeown2016} is also in fair agreement with weak antilocalization and quantum oscillation measurements and calculations of interface 2DEGs~\cite{Caviglia2010,Fete2014,Hurand2015,Liang2015}. Moreover, the nature and strength of electron-phonon interaction at both the bare STO (001) surface and the LAO/STO (001) interface were found to be in good agreement~\cite{Wang2016,Boschker2015,Cancellieri2015}. 
The universality of electronic properties is further supported by a recent experiment demonstrating that room temperature deposition of $\approx2$~\AA{} Al on a \tio{} terminated STO (001) wafer surface donates electrons and induces a 2DEG in STO \cite{Rodel2016} by strongly reducing the surface. As seen in \figref{Rodel2016_F1} the subband structure of this 2DEG bears all the hall marks of the STO surface 2DEG, while in fact it exists at the interface of STO and aluminium oxide.

On the other hand, some basic 2DEG properties such as the carrier density and occupied bandwidth remain controversial. For the surface 2DEGs discussed in this chapter, the latter is evident from the ARPES data while the total carrier concentration and density of states at the Fermi level are difficult to determine experimentally since the closely spaced shallow subbands predicted by band structure calculations have so far eluded detection. In interface systems, both carrier density and bandwidth are difficult to quantify. The Hall effect, commonly used as a measure of the carrier density, is strongly non-linear for the LAO/STO interface and its quantitative relation to the itinerant carrier density is model dependent~\cite{FeteThesis}. Furthermore estimates of the occupied bandwidth based on quantum oscillation and spectroscopic measurements range from $\approx 5$~meV~\cite{McCollam2014} up to $\approx 200$~meV~\cite{Berner2010,Boschker2015} for LAO/STO systems with similar Hall densities. Thus so far it has proved difficult engineer these fundamental properties through control of the growth conditions. A direct comparison of results from ARPES and transport experiments on a single STO 2DEG sample could provide more insight and help to establish the extent to which electronic properties can be considered universal. However, this remains a challenging task as the surface 2DEG is not accessible to standard magneto-transport experiments while ARPES experiments on interfaces are limited to the soft X-ray regime where the effective resolution has so far precluded results that resolve the full subband structure.
In this regard, the aluminium oxide/STO interface studied by ARPES in Ref.~\cite{Rodel2016} provides an interesting opportunity to overcome these difficulties as it is much more amenable to magneto-transport experiments than the bare STO surface due to the protective nature of the aluminium oxide over-layer.

Another aspect that has perhaps not received the attention it deserves is the relation between sample environment and 2DEG properties. In section~\ref{OV_section} we described the exceptional sensitivity of the cleaved STO (001) surface to light induced oxygen vacancy formation as well as to the residual oxygen partial pressure. Adsorbates on the LAO surface and irradiation even with visible light were also found to strongly affect the LAO/STO interface 2DEG~\cite{Tebano2012,Scheiderer2015,Brown2016}. Yet, the understanding of the effects of different sample preparations and environments including electron and photon beams, as they are used in many experiments, is only in its infancy and much work remains to be done to improve the consistency of experiments and ultimately the stability of devices. On the other hand, the sensitivity of emergent properties in oxides to defects, including those introduced by experimental probes, offers entirely new perspectives for tailoring properties on the nanoscale~\cite{Cen2008,Cheng2015,Meevasana2011a}.

\begin{acknowledgement}
The authors acknowledge financial support through the University of Geneva and the Swiss National Science Foundation and would like to thank C. Bernhard, R. Claessen, A. F\^ete, S. Gariglio, K. Held, P.D.C. King, F. Lechermann, L.D. Marks, W. Meevasana, N.C. Plumb, M. Radovic, V. Strocov, A. Tamai, J.M. Triscone, D. van der Marel and Z. Wang for discussions.
\end{acknowledgement}

%\input{referenc}
%\printbibliography[heading=bibintoc]	% print bibliography here
\bibliographystyle{plain}%abbrv}
\bibliography{Chapter.bib}%,FB.bib}
\end{document}